\documentclass[twocolumn,superscriptaddress,footinbib,amsmath,amssymb,aps,prb,letterpaper,10pt,longbibliography,floatfix]{revtex4-2}

\usepackage{graphicx}
\usepackage{dcolumn}
\usepackage{soul}
\usepackage{bm,physics}
\usepackage[utf8]{inputenc}
\usepackage[english]{babel}
\usepackage[dvipsnames]{xcolor}
\usepackage[colorlinks=true, urlcolor=blue, pdfborder={0 0 0}]{hyperref}
\usepackage{mathtools}
\usepackage{siunitx}
\usepackage{tikz}
\usetikzlibrary{decorations.markings,decorations.pathreplacing,arrows,arrows.meta,calc,shapes.arrows,shapes.geometric}
\usepackage[caption=false]{subfig}
\usepackage{verbatim}

\newcommand{\req}[1]{Eq.~(\ref{#1})}
\newcommand{\reqs}[1]{Eqs.~(\ref{#1})}
\newcommand{\rref}[1]{(\ref{#1})}

\begin{document}

	\title{Mobile Topological Su-Schrieffer-Heeger Soliton in a  Josephson Metamaterial}
	\author{Dushko Kuzmanovski}
	\affiliation{
		Nordita, KTH Royal Institute of Technology and Stockholm University
		Hannes Alfv\'{e}ns v\"{a}g 12, SE-106 91 Stockholm, Sweden
	}
	\author{Rub\'{e}n Seoane Souto}
     \affiliation{Instituto de Ciencia de Materiales de Madrid (ICMM), Consejo Superior de Investigaciones Cient\'{i}ficas (CSIC), Sor Juana In\'{e}s de la Cruz 3, 28049 Madrid, Spain}
	
	\author{Patrick J. Wong }
\affiliation{
		Nordita, KTH Royal Institute of Technology and Stockholm University
		Hannes Alfv\'{e}ns v\"{a}g 12, SE-106 91 Stockholm, Sweden
	}
	\affiliation{
		Department of Physics, University of Connecticut, Storrs, Connecticut 06269, USA
	}
 
	\author{Alexander V. Balatsky} 
	\affiliation{
		Nordita, KTH Royal Institute of Technology and Stockholm University
		Hannes Alfv\'{e}ns v\"{a}g 12, SE-106 91 Stockholm, Sweden
	}
	\affiliation{
		Department of Physics, University of Connecticut, Storrs, Connecticut 06269, USA
	}
	\date{\today}
	
	\begin{abstract}
    Circuits involving arrays of Josephson junctions have emerged as a new platform for exploring and simulating complex bosonic systems. Motivated by this advance, we develop and theoretically analyze a one-dimensional bosonic system with sublattice symmetry, a bosonic Su-Schrieffer-Heeger model. The system features electrostatically controlled topological mid-gap states that we call soliton states. These modes can be measured using either spectroscopy through a normal lead or admittance measurements. We develop a protocol
to adiabatically shuttle the position of these topological soliton states using local electrostatic gates. We
demonstrate a nearly perfect fidelity of soliton shuttling for timescales within experimental reach.
	\end{abstract}
	
	\maketitle

\section{\label{sec:Intro}Introduction} 
A recent paradigm in quantum technology is the advent of quantum computation. The hardware of a general-purpose quantum computer which has the ability to generically run any quantum algorithm~\cite{NeillNat2021} has not yet been realized.
A complementary approach is utilizing special purpose-built hardware~\cite {GeorgescuRMP2014} for simulating specific many-body Hamiltonians. Platforms based on ion traps~\cite{MonroeRMP2021}, cold atoms~\cite{Bloch2012},  and circuits involving Josephson junctions~\cite{Schmidt2013} have been used for this purpose. Superconducting platforms show advantages regarding scalability, low power consumption, low noise, and ease of interfacing. These systems involve several islands that can exchange Cooper pairs via tunneling. Fabrication advances have enabled islands to be defined with sub-micron precision, allowing the investigation of superconductor-insulator transition~\cite{GeerligsPRL1989,ChenPRB1995,ZantPRB1996,BaskoPRB2020}, including anomalous metal phases~\cite{BoettcherNat2018}. Apart from the current interest in quantum information processing applications, quantum metamaterials based on Josephson-junction arrays offer the possibility of creating platforms for investigating the effects of circuit quantum electrodynamics~\cite{RakhmanovPRB2008, MachaNatComm2014, BrehmAPL2022}.

In this work, we theoretically analyze the formation of topological states in a chain of superconducting (SC) islands connected via Josephson junctions forming a Josephson junction array (JJA), Fig~\ref{fig:jjalithograph}. We work in a regime where junction shunt capacitances are negligible relative to gate capacitances. We consider that the Josephson energy between the islands has an amplitude that alternates, realizing a superconducting version of the Su-Schrieffer-Heeger (SSH) model. We demonstrate that local electrostatic gates can tune the position of these states, which can be detected by either tunneling or noise spectroscopy. For instance, detuning the energy of one island away from zero, which can be seen as adding a defect into the system, creates a topological state.

The SSH model is a paradigmatic topological insulator in one dimension~\cite{su1979,heeger1988,asboth2016}. It is characterized as a tight-binding chain of 2-site unit-cell dimers with an intracell hopping $t_1$ and an intercell hopping $t_2$. For a given trermination of the chain, the model is either in a trivial insulator phase with a gapped spectrum, or in a topological phase with zero-energy states localized on its boundaries, depending  on the ratio of the hopping elements, $t_2/t_1 < 1$ or $t_2/t_1 > 1$. The SSH model hosts localized states near domain walls, where the parity of the dimerization changes~\cite{asboth2016}. These localized in-gap states correspond to soliton states in the continuum theory of the SSH model~\cite{takayama1980} and we adopt this nomenclature in our discussion.

The bosonic SSH model has been realized in many other systems, such as optical waveguides~\cite{chen2021}, acoustic phonons~\cite{liu2020}, magnons~\cite{mei2019,li2021,WeiJPCM2022}, Rydberg atoms~\cite{meier2016,deleseleuc2019}, ultracold atoms~\cite{atala2013,lohse2016,xie2019}, and superconducting transmons~\cite{mei2018,nie2020,guan2023interplay}. Furthermore, a similar concept to ours was considered where the elementary excitations were fluxons~\cite{PetrescuPRB2018}.

We develop a protocol to shuttle topological states along the chain by tuning the energy of adjacent islands. Topological states have orthogonal wavefunctions. Therefore, detuning additional sites would result in an exact crossing of the wavefunctions and a failure of the protocol. Instead, bringing the energy of the impurity island close to resonance, makes the ground state hybridize with the continuum. If the process is done adiabatically, there is no weight transferred to the continuum of states. In this way, there is a finite overlap of the wavefunction with the one where the soliton state is located at another position. The fast and efficient manipulation of soliton defect states can be utilized for implementing robust mesoscopic qubits, as well as racetrack-type quantum memories.~\cite{ParkinIEEE2003, ParkinScience2008, FertNatNano2013, ParkinNatNano2015}

The rest of this paper is organized as follows: In Sec.~\ref{sec:Model} we introduce the Hamiltonian governing a general JJA circuit and describe the parameter regime in which we obtain an effective tight-binding model. In Sec.~\ref{sec:Results}, we focus on a linear SSH chain with applied on-site potential as control of a soliton state. The main conclusions are given in Sec.~\ref{sec:Conclusion}. Additionally, we demonstrate an alternative shuttling protocol with low fidelity in Appendix \ref{App:Protocol2}, and an animation of the two shuttling protocols over a whole ring in Appendix~\ref{App:Animation}.

\section{Model\label{sec:Model}}
We model the physical realization of an array of Josephson junctions, Fig~\ref{fig:jjalithograph} by considering the circuit depicted in Fig.~\ref{fig:jjaschematic}. We write down a circuit Hamiltonian following the circuit theory of Refs.~\cite{Vool_IJCTA2017, FazioPhysRep2001} as
	\begin{equation}
 \begin{multlined}
		H_{\mathrm{JJA}} = \frac{1}{2} \, \sum_{i, j} \tilde{C}^{-1}_{i, j} \, \qty(n_{i} - n_{g, i}) \, \qty(n_{j} - n_{g, j}) \\
		- \sum_{\langle i, j \rangle} J_{\langle i, j \rangle} \, \cos\qty(\varphi_{i} - \varphi_{j}), \label{eq:CircuitHam1}
  \end{multlined}
	\end{equation}
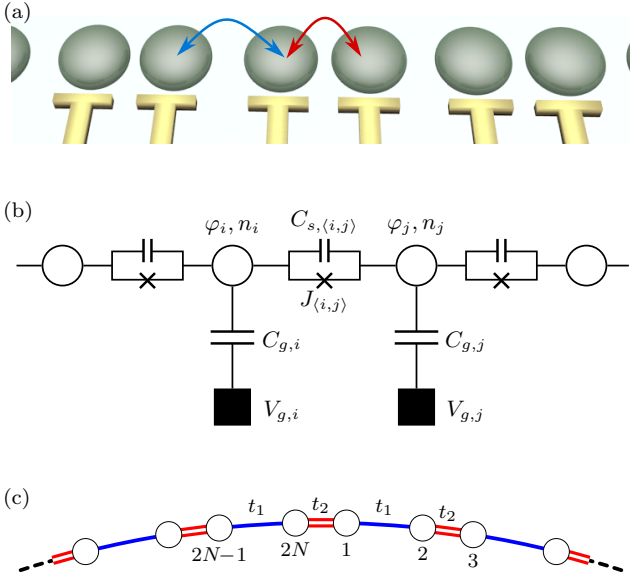
\begin{figure}
    \begin{tikzpicture}
        \node (fig) at (0,0) {\includegraphics[width=0.95\columnwidth]{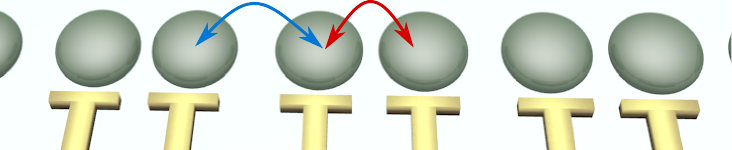}};
        \node at (fig.north west) {\subfloat[\label{fig:jjalithograph}]{}};
    \end{tikzpicture}
    \\
    \begin{tikzpicture}
        \node (fig) at (0,0) {\includegraphics[width=0.95\columnwidth]{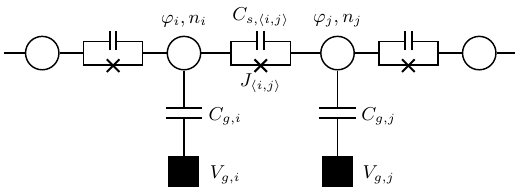}};
        \node at (fig.north west) {\subfloat[\label{fig:jjaschematic}]{}};
    \end{tikzpicture}
    \\
    \begin{tikzpicture}
        \node (fig) at (0,0) {\includegraphics[width=0.95\columnwidth]{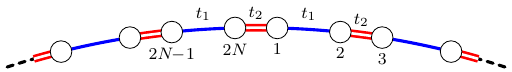}};
        \node at (fig.north west) {\subfloat[\label{fig:ssh_ring}]{}};
    \end{tikzpicture}
    \caption{\label{fig:sketch}Three representative schematics of our Josephson junction array (JJA).
    	\protect\subref{fig:jjalithograph} Representation of the chain of superconducting (SC) islands and voltage gates. Tunneling of Cooper pairs between nearby islands have alternating strengths symbolized by the blue and red arrows. \protect\subref{fig:jjaschematic} Circuit diagram of SC islands parameterized by $\varphi$ and $n$ coupled via JJs with critical currents $J_{\langle i, j\rangle}$ and shunt capacitances $C_{s,\langle i, j\rangle}$. Each island is capacitatively coupled via a gate capacitance $C_{g, i}$ to an externally voltage-controlled gate $V_{g, i}$. \protect\subref{fig:ssh_ring} Effective SSH tight-binding model of a JJA in ring geometry with $N$ dimers. The blue links are ``weak'' hopping amplitudes $t_{1}$, while the red double links are ``strong'' hoppings $t_{2}$. On-site potentials can be applied at each  particular site.
    }
\end{figure}
	where the mutual capacitance matrix is
	\begin{equation}
		\tilde{C}_{i, j} = \qty(C_{g, i} + \sum_{k} C_{s, \langle i, k \rangle}) \, \delta_{i, j} - C_{s, \langle i, j\rangle}, \label{eq:CapMat}
	\end{equation}
	while the charge offsets $n_{g, i}$ due to applied gate voltages $V_{g, i}$ are
	\begin{equation}
		n_{g, i} = -C_{g, i} \, V_{g, i}. \label{eq:ChargeOffSet}
	\end{equation}
	The gate voltages are in principle time-dependent and serve to control the charge offset of every SC island. We have adopted a system of units in which $\hbar = 2 e = 1$. In this case, the Josephson critical current and Josephson energy have the same units, and the inverse capacitance matrix equals the charging energy. The canonical commutation relation of the conjugate variables is
	\begin{equation}
		\qty[\varphi_{i}, n_{j}] = i \, \delta_{i, j}, \label{eq:CanonCommRel}
	\end{equation}
	with all the other commutators being identically zero. One can go over from phase and number variables to canonical bosonic creation-annihilation operators
	\begin{equation}
		b^{\dagger}_{i} = n^{\frac{1}{2}}_{i} \, e^{i \, \varphi_{i}}. \label{eq:Creation1}
	\end{equation}
    With this in mind, the Josephson term in \req{eq:CircuitHam1} is reminiscent of a Cooper-pair hopping term, but includes interactions due to the non-linearity in \req{eq:Creation1}. At the same time, the charging energy term in \req{eq:CircuitHam1} is a four-point interaction term in terms of the creation/annihilation operators. We can approximate the Hamiltonian to a non-interacting one in the limit where charging energy dominates, so the number of Cooper pairs can only fluctuate between $n_i$ and $n_i+1$ ($n_{i}$ being an integer)
 	~\footnote{The mean-field contribution of the charging energy is an extra tight-binding element $\sum_{i, j} m_{i, j} \, b^{\dagger}_{i} \, b_{j}$, where $m_{i, j} = m^{\ast}_{j, i}$ is determined self-consistently
	\[
		m_{i, j} = \tilde{C}^{-1}_{i, j} \, \expval{b^{\dagger}_{j} \, b_{i}}.
	\]
	The assumption is that this matrix is a small correction to the hopping elements already included in \req{eq:HopHam1}.
	}.
 In this way, we arrive at the following non-interacting Cooper-pair hopping tight-binding Hamiltonian
	\begin{eqnarray}
		& \mathcal{H}_{\mathrm{CPH}} = -\sum_{i, j} t_{i, j} \, b^{\dagger}_{i} \, b_{j} + \sum_{i} V_{i} \, n_{i}, \label{eq:HopHam1}
	\end{eqnarray}
	where the effective hopping element is
	\begin{equation}
		t_{i, j} = \frac{J_{\langle i, j \rangle}}{2 \sqrt{\expval{n_{i}} \, \expval{n_{j}}}}, \label{eq:HopTerm}
	\end{equation}
	and the effective on-site potential is
	\begin{equation}
		V_{i} = \sum_{j} \tilde{C}^{-1}_{i, j} \qty(\expval{n_{j}} - n_{g, j}). \label{eq:OnSiteTerm}
	\end{equation}

	The excitations of \req{eq:HopHam1} are described by linear combinations of the on-site creation operators
	\begin{equation}
		\Gamma^{\dagger}_{\alpha} = \sum_{i} u^{\qty(\alpha)}_{i} \, b^{\dagger}_{i},
	\end{equation}
	where the effective single-particle wavefunctions $u^{\qty(\alpha)}_{i}$ of the $\alpha$-th excitation satisfy the eigenvalue problem
	\begin{equation}
		-\sum_{j} t_{i, j} \, u^{\qty(\alpha)}_{j}  + V_{i} \, u^{(\alpha)}_{i} = \varepsilon_{\alpha} \, u^{\qty(\alpha)}_{i}. \label{eq:WaveEqn}
	\end{equation}
	and the normalization condition
	\begin{equation}
		\sum_{i} \abs{u^{(\alpha)}_{i}}^{2} = 1. \label{eq:Norm1}
	\end{equation}
 Then, each of these modes is occupied according to a Bose-Einstein distribution
	\begin{equation}
		\expval{\Gamma^{\dagger}_{\alpha} \, \Gamma_{\beta}} = n_{B}\qty(\varepsilon_{\alpha}/T) \, \delta_{\alpha, \beta}. \label{eq:ExcExpVal}
\end{equation}
where $n_{B}\qty(x) = 1/\qty(\exp\qty(x)- 1)$, and $T$ is an effective temperature of the Cooper-pair subsystem. This is achieved by tuning a uniform shift of each $V_{i}$ in \req{eq:OnSiteTerm} by a sufficiently large positive value, which, accordingly, shifts each excitation energy $\varepsilon_{\alpha}$, so that the total number of Cooper pairs in the JJA, $N_{\mathrm{CP}}$, satisfies
	\begin{equation}
		\expval{N_{\mathrm{CP}}} = \sum_{\alpha} n_{B}\qty(\varepsilon_{\alpha}/T). \label{eq:TotNumCP}
	\end{equation}
	The site occupancy that enters in \reqs{eq:HopTerm}, \rref{eq:OnSiteTerm} is given by
	\begin{equation}
		\expval{n_{i}} = \sum_{\alpha} \abs{u^{\qty(\alpha)}_{i}}^{2} \, n_{B}\qty(\varepsilon_{\alpha}/T). \label{eq:SiteOcc}
	\end{equation}

\section{\label{sec:Results}Results}
The discussion in Sec.~\ref{sec:Model} is general and applies for any JJAs where charging energy dominates. We now focus on simulating the SSH model. To avoid finite-size effects, we consider a ring geometry with alternating hopping elements as indicated in Fig.~\ref{fig:ssh_ring}. In Sec.~\ref{subsec:Soliton}, we demonstrate how to introduce topological defects via applying on-site control potential. In Sec.~\ref{subsec:Berry}, we give several descriptions of the topological nature of these soliton states bound to the control sites. Sec.~\ref{subsec:Admittance} gives a concrete experimentally measurable signature of the soliton state, and in Sec.~\ref{subsec:Shuttle} we discuss a protocol for shuttling this soliton state to the neighboring dimer.

\subsection{\label{subsec:Soliton}Electrostatically controlled soliton states}	
Here, we explore the topological states appearing in the system when creating a defect using on-site potential. This has the advantage that the position of the defect state is controllable with external voltages, whose time scales are, in principle, much faster than any electrostatic manipulation of the critical current in the JJ links.
		\begin{figure}[!ht]
				\includegraphics[scale=1.0]{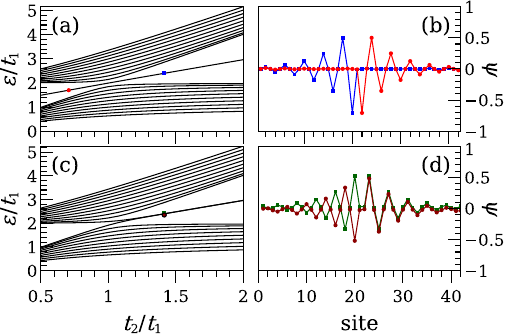}
				\caption{\label{fig:BoundState} Spectrum (panels (a) and (c)) on a ring with $N = 21$ dimers as a function of the ratio of hopping terms $t_{2}/t_{1}$. Panels (b) and (d) show the in-gap bound state real wavefunctions for the corresponding value of the parameter highlighted by dots in the left panels. The chemical potential is adjusted so that the average occupancy per site is $\expval{n} = 5.0$, and the effective temperature is always $\textit{T}/\textit{t}_{1} = 0.0005$ for large on-site potential turned on at site $21$ (panels (a) and (b)), and turned on at sites $21$ and $22$ (panels (c) and (d)).}
		\end{figure}

    We begin by discussing the effects of detuning one island (or an odd number of consecutive islands) away from resonance. For a sufficiently large detuning of the island, there is always an in-gap bound state, regardless of the ratio of bond hopping elements $t_{2}/t_{1}$, as depicted in Fig.~\ref{fig:BoundState}(a). The only difference is the direction towards which the wavefunction of such a state spreads: the soliton extends toward the direction of the strong broken bond. As the ratio between the hopping changes from $t_{2}/t_{1}<1$ to $t_{2}/t_{1}>1$, the position of the nearest-neighbor bond to the impurity site with larger hopping changes from right to left, and, with it, the orientation of the wavefunction spread, as shown on Fig.~\ref{fig:BoundState}(b) for two representative values of the hopping-elements ratio. The behavior of the soliton wavefunction suggests that turning on a potential on another site would have the most significant detrimental effect due to destructive interference if it is on a neighboring site connected via a strong bond to the first potential site. Conversely, if connected via a weak bond, it would not hinder the spread of the SSH soliton wavefunction.
    
    We now consider the effect of turning on a strong repulsive on-site potential on a dimer. Applying such a control potential profile has the same effect as removing a dimer (control dimer) and creating two artificial edges on the sites closest to the control dimer in the ring geometry. If the control dimer is linked to the rest of the chain via strong bonds, topological SSH states appear at the two sides of the of the dimer. No states appear in the case this links are weak.
    
    Figure~\ref{fig:BoundState}(c) shows the energy spectrum as a function of $t_{2}/t_{1}$ for a ring with detuned dimer that connects via $t_2$ (and  connected via $t_1$ to the rest of the ring). When the ratio goes from less to above one, two topological states detach from the continuum and pin at the center of the gap. The wavefunctions of these two nearly-degenerate states are depicted in Fig.~\ref{fig:BoundState}(d).
    
    Since the soliton state of an SSH model is localized at a domain wall, the soliton can move by shifting the domain wall. Conventionally, this is accomplished by modulating the hopping parameters. However, instead of modifying the ratio $t_{2}/t_{1}$, we can achieve an effective interchange of $t_{1}$ and $t_{2}$ by simply shifting the two positions of the on-site potentials by one site. We illustrate this possibility in the two panels in Fig.~\ref{fig:Admittance}. In Fig.~\ref{fig:Admittance}(a), dimer bond is strong, thus two mid-gap solitons, and in Fig.~\ref{fig:Admittance}(b) the bond is weak, and no mid-gap states are present.

 \subsection{\label{subsec:Berry}Topological properties of the soliton state}
 In this subsection we wish to establish quantitatively the topological nature of the soliton states produced by introducing a potential on the control dimer. The localized wave function of band $\alpha$ takes the asymptotic form of $\left|u^{(\alpha)}_{2j}\right|^2 = \left|u^{(\alpha)}_{j_0}\right|^2 e^{-2j/\xi^{(\alpha)}}$ for site $2j$ in the limit of large system size where $j_0$ is the site hosting the localized soliton state and $\xi$ is a localization length. We take the $2j$-th site as the localized wave functions only have support on one sublattice. The localization length of the SSH model is $\xi_{\mathrm{SSH}} = \ln(t_2/t_1)$~\cite{asboth2016} where in the topological phase $t_2/t_1>1$. For the parameters chosen here ($t_2=2t_1$), $\xi_{\mathrm{SSH}} = \ln(2) = \num{0.6931}\ldots$. For the wave function on the JJA ring with $V/t_1 = 100.0$ on the control dimer, we find that $\xi^{(20)}_{\mathrm{JJA}} = \num{0.6925}(1)$ and $\xi^{(21)}_{\text{JJA}} = \num{0.6932}(1)$, in close agreement with the topological SSH value. The exact SSH value is valid for states with exactly zero energy, which is obtained in the limit of $V/t_1 \to \infty$, and where the wave function has support on one sublattice only, which is valid in the case of $N\to\infty$. We find that $|\xi_{\mathrm{JJA}} - \xi_{\mathrm{SSH}}| \sim 1/V$. This reinforces our identification of the localized state as being topological in origin.

 For a one-dimensional system, its topology can be characterized by the winding number, or Zak phase~\cite{zak1989}. A winding number can be calculated in the conventional way by taking the entire ring geometry to be representative of a periodic unit-cell and then proceeding to construct a momentum space representation. This is conventionally performed for the standard 2-sites per unit cell SSH model, however it has also been performed for the 4-site per unit cell SSH$_4$ model \cite{Eliashvili2017,Maffei2018}, as well as arbitrarily large unit cell extensions of the SSH model~\cite{wong2022}.
 The Zak phase of a band $n$ is defined as
 \begin{equation}
    \gamma_n = \frac{1}{\pi} \int_{0}^{2\pi} dk A_n(k) ,
 \end{equation}
 where $A_n(k) = i \psi_n^\dagger(k) \partial_k \psi_n(k)$ is the Berry potential of the $n$-th band with $\psi_n(k)$ the eigenvector for the $n$-th band of the momentum space Hamiltonian in the chiral basis.
 In the chiral basis the Hamiltonian takes the form of
 \begin{equation}
    H(k) = \begin{pmatrix} \epsilon & h(k) \\ h^\dagger(k) & \epsilon \end{pmatrix}
 \end{equation}
 where
 \begin{align}
    h(k) &=
    \begin{pmatrix}
        t_1 & t_2 & 0 & \cdots & t_2 e^{i k}
        \\
        0 & t_1 & t_2 & \ddots & 0
        \\
        \vdots & \ddots & \ddots & \ddots & \vdots
        \\
        \vdots & \ddots & \ddots & t_1 & t_2
        \\
        0 & \cdots & \cdots & 0 & t_1
    \end{pmatrix}
 \end{align}
 and
 \begin{align}
    \epsilon &= \begin{pmatrix} V & 0 & \cdots \\ 0 & 0 & \ddots \\ \vdots & \ddots & \ddots \end{pmatrix} .
 \end{align}
 The chiral basis is defined as the basis in which the Hamiltonian takes the form of
 \begin{equation}
    \mathcal{H}_{\mathrm{CPH}} = \chi^\dagger H \chi
 \end{equation}
 where $\chi = (b_1,b_3,\ldots,b_{2N-1},b_2,b_4,\ldots,b_{2N})^T$.
 The winding number can be calculated numerically using a method adapted from the scheme of Ref.~\cite{fukui2005} for numerical evaluation of the Chern number. For numerical evaluation the Brillouin zone is discretized $k \mapsto k_a$ where $a$ is a discrete index with the $k_a$ spaced by the interval $\Delta k = |k_{a+1} - k_{a}|$. The Zak phase can be calculated using this discretization as
 \begin{equation}
 \begin{aligned}[b]
    e^{-i \gamma_n}
    &=  e^{-i \oint A_n(k) dk}
    \\
    &\approx e^{-i \sum_a A_n(k_a) \Delta k}
    \approx \prod_a \left[ \psi_n^\dagger(k_a) \psi_n(k_{a+1}) \right]
 \end{aligned}
 \end{equation}
 where we have employed the unit normalization of the wavefunction and the discrete derivative.
 The phase is then obtained by taking the principal value of the logarithm,
 \begin{equation}
    \gamma_n = \frac{i}{\pi} \ln\left( \prod_{a} \psi_n^\dagger(k_a) \psi_n(k_{a+1}) \right) .
 \end{equation}
 The product is taken over the Brillouin zone, $k_a \in [0,2\pi]$.
 We choose to index the unit-cell such that the dimer hosting the control potentials is located on the first dimer of the unit cell.
In the topological phase, $t_2 / t_1 > 1$, the zero energy bands have winding numbers 
 \begin{align}
    \gamma_{20} &= -1 ,& \gamma_{21} &= +1
\intertext{and in the trivial phase, $t_2 / t_1 < 1$, have winding numbers}
    \gamma_{20} &= 0 ,& \gamma_{21} &= 0 .
\end{align}
This calculation demonstrates that it is appropriate to identify the soliton states as being topological.

\subsection{\label{subsec:Admittance}Signatures of soliton state in the on-site admittance spectrum}
	\begin{figure}[!ht]
		\includegraphics[scale=0.9]{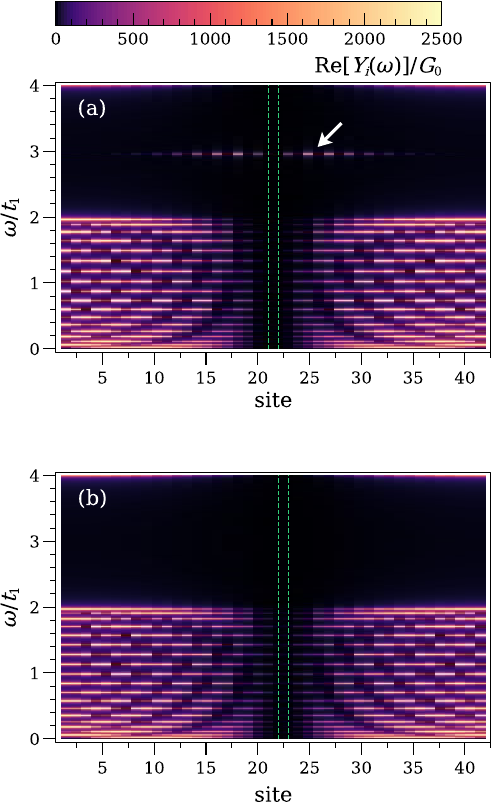}
		\caption{\label{fig:Admittance}Real part of local admittance $\Re\qty[Y_{i}\qty(\omega)]$ as a function of site $i$ and frequency $\omega$. (a) On-site potential applied on dimer $21$-$22$ with a weak bond between the control sites. The white arrow indicates the peaks associated with transition to the soliton dimer states. (b) On-site potential applied on dimer $22$-$23$ with a strong bond between the control sites. The ratio of the hopping elements is chosen to be $t_{2}/t_{1} = 2.0$, and the applied control potential is $V/t_{1} = 100.0$. We choose to work at temperature $T/t_{1} = 0.005$ and the uniform potential shift is $V/t_{1} = 2.992$ in both panels so that the average occupancy per site is $\expval{n} = 5.0$. The smearing factor in the Lorentzian representation of a delta function $\delta\qty(x) \approx \eta/\qty[\pi \, \qty(x^{2} + \eta^{2})]$ is chosen to be $\eta = 0.005$. The scaling function chosen for scaling the color coding is $I(G) = 1 - e^{-\frac{G}{\delta G}}$, with $\delta G/G_{0} = 25$, where $G_{0} = 2 e^{2}/h$.}
	\end{figure}

To elucidate the electronic response of the devise hosting SSH state we consider admittance. According to the fluctuation-dissipation theorem~\cite{SoutoPRR2020}, the admittance is related to the number-number spectral density function $\rho_{n_{i}, n_{j}}\qty(\omega)$, with the included modification provided by the capacitative matrix. A signature of this state is visible in the finite-frequency spectrum probed locally by measuring the site-dependent AC admittance via small-signal analysis on the same gates where the polarizing gate potential is applied. 
    
    The exact expression is
    
    \begin{equation}
		\Re\qty[Y_{i}\qty(\omega)] = \omega \, C^{2}_{g, i} \, \sum_{k, l} \tilde{C}^{-1}_{i, k} \rho_{n_{k}, n_{l}}\qty(\omega) \, \tilde{C}^{-1}_{j, l}. \label{eq:ReAdmit}
	\end{equation}
 
    Working in the approximation of a hopping Hamiltonian Eq.~\eqref{eq:HopHam1}, $\rho_{n_{i}, n_{j}}\qty(\omega)$ is given by the following expression involving the eigenvalues and wavefunctions
    
    \begin{equation}
    \begin{multlined}
		\rho_{n_{k}, n_{l}}\qty(\omega) = \pi \, \sum_{\alpha, \beta} u^{\qty(\alpha)}_{k} \, \qty(u^{\qty(\alpha)}_{l})^{\ast} \, u^{\qty(\beta)}_{l} \, \qty(u^{\qty(\beta)}_{k})^{\ast} \times \\
		\times \qty[n_{B}\qty(\tfrac{\varepsilon_{\beta}}{T}) - n_{B}\qty(\tfrac{\varepsilon_{\alpha}}{T})]
		 \, \delta\qty(\omega - \varepsilon_{\alpha} + \varepsilon_{\beta}). 
   \end{multlined}
   \label{eq:SpecDensityNumber}
	\end{equation}

	\begin{figure*}[!ht]
		\includegraphics[scale=0.9]{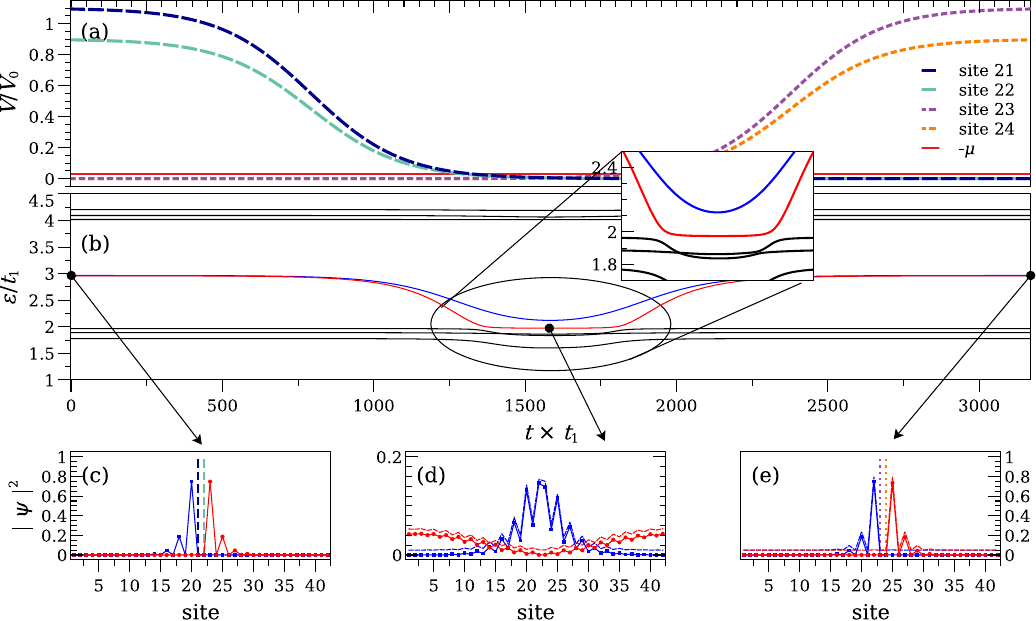}
		\caption{\label{fig:Protocol1}Adiabatic protocol P1 for shuttling the soliton dimer states from dimer $21$-$22$ to dimer $23$-$24$. (a) Profile of the ramp dimer potential in the course of the time steps. The degeneracy is broken by taking $1.1 \, V_{0}$ on the left site and $0.9 \, V_{0}$ on the right site of each dimer. Note that there is a uniform chemical potential applied to each site to keep the average number of particles fixed as in the previous figures. The time step $\Delta t = \pi/(1.1 \times 2 \times  \qty(\abs{t_{1}}+ \abs{t_{2}}))$ is small enough so that the Nyquist frequency is higher than the bulk bandwidth. (b) The energies of the  instantaneous eigenfunctions of the time-dependent Hamiltonian corresponding to the on-site potential from panel (a). Panels (c), (d), and (e): Snapshots of the time evolution of the single-particle wavefunction initialized in the corresponding initial eigenstate  with the same color coding as in panel (b). The vertical dashed lines indicate the bond where the on-site potential is applied. The dashed curve on panel (e) is the instantaneous eigenstate at that time instant for visual comparison. The other parameters are identical as in Figs.~\ref{fig:BoundState}, \ref{fig:Admittance}. In panels (d), (e), the dashed line wavefunctions are the instantaneous eigenfunctions shifted by a tiny amount for visibility.}
	\end{figure*}

Independently from the regime, the admittance is suppressed at the control dimer, due to the high detuning potential considered, Fig.~\ref{fig:Admittance}. We note a quasi-continuum spectrum coming from the states above the gap. We note a discrete transition in Fig.~\ref{fig:Admittance}(a) coming from the in-gap states, that is absent in the case that the system does not feature sub-gap states , Fig.~\ref{fig:Admittance}(b). The  admittance signal of the state oscillates in space, decaying eventually, being a measurement of the localization properties of the sub-gap state.

Alternatively, local spectroscopy of the the system using weakly-coupled metallic leads can reveal a similar information. We expect a peak in the conductance when the leads bias voltage align with the energy of the state. However, metallic probes can be detrimental for the shuttling protocol below, as they can induce uncontrolled changes between the two ground states.

\subsection{\label{subsec:Shuttle}Shuttling protocol the SSH soliton state}
\begin{figure}[!ht]
        \centering
        \includegraphics[scale=0.9]{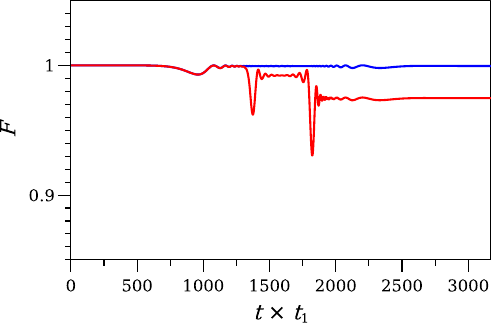}
        \caption{\label{fig:FidelityP1} Fidelity for a shuttling process between neighboring dimers, \req{eq:FidelityDef} for the two mid-gap soliton states in Fig.~\ref{fig:Protocol1} during protocol P1. Note the small range of values on the vertical axis.}
    \end{figure}
Now, we consider shuttling the soliton state by a single dimer via an adiabatic sweep of the potential controlling two neighboring islands. The coherence of the soliton state during the shuttling is dependent upon the protocol used for the potential sweep. The procedure which successfully maintains the state is illustrated in Fig.~\ref{fig:Protocol1} and is referred to as protocol P1. For protocol P1, we initially consider large potentials on two adjacent sites belonging to the same cell. We focus on the topological regime ($t_2/t_1>1$), where the system shows in-gap states. The potential is slowly ramped down on these sites from a largely detuned situation to be on resonance. At the same time, we ramp up the potential of a neighboring cell. In order to break the degeneracy of the two solition states (Fig.~\ref{fig:BoundState}(c)), we choose a slightly different value ($\pm 10 \, \%$) on the two control sites of the dimer, see Fig.~\ref{fig:Protocol1}(a). Figs.~\ref{fig:Protocol1}(c), (d), (e) depict the wavefunctions of several instantaneous eigenvalues around the mid-gap states during the shuttling protocol. The vertical dashed lines show the control dimer having strong on-site potential in Fig.~\ref{fig:Protocol1}(c), (e). We note a significant overlap of the time-evolved wavefunction with the final one, represented by the thin dashed line. We see that the mid-gap states always avoid crossing with the top of the valence band for a finite chain length, a  necessary constraint for preserving the information encoded in them. They remain nearly degenerate until the on-site potential is about $4$ times the bulk bandwidth, apart from the degeneracy point when the on-site potential is turned off in the middle of the protocol.

As a more quantitative measure of the protocol's success, in Fig.~\ref{fig:FidelityP1} we illustrate the overlap integral of the time-evolved state $U\qty(t) \, \ket{u^{\qty(\alpha)}(0)}$ with the instantaneous eigenstate $\ket{u^{\qty(\beta)}(t)}$
\begin{equation}
    \label{eq:FidelityDef}
    F_{\alpha}\qty(t) = \abs{\mel{u^{\qty(\alpha)}\qty(t)}{U\qty(t)}{u^{\qty(\alpha)}\qty(0)}}^{2}, \ \alpha \in \Bqty{s_{1}, s_{2}}
\end{equation}
for the two soliton states $s_{1}$, $s_{2}$, which have color coding red and blue, respectively in Fig.~\ref{fig:Protocol1}.

We note the remarkably high fidelity level during this protocol that can be traced to the finite energy gap between the mid-gap states and the top of the valence band (Fig.~\ref{fig:Protocol1}(b)) throughout the whole procedure. The avoided crossing with the states at the quasi-continuum is smaller in the case of the lower-lying state (red). That results in a lower fidelity with respect to the higher state. In any case, it can be optimized to be arbitrarily close to $F=1$ using slower sweep rates. We note the importance of breaking the degeneracy between the two sites of the dimer. In its absence, there is a significant conversion from one subgap state into the other, that can reach values up to $10 \, \%$ to $20 \, \%$. This weight transference takes pace at the middle of the process, where the states hybridize with the continuum.

As an illustration of the importance of the relative time shift between the ramp-down and ramp-up phases in Fig.~\ref{fig:Protocol1}(a), we illustrate an alternative protocol P2 in Appendix~\ref{App:Protocol2}. In this case, we start the ramp-up well before the solitons hybridize with the continuum. Due to the orthogonality of the wavefunctions, the protocol fails and $F\to0$, Fig.~\ref{fig:FidelityP2}.

\section{\label{sec:Conclusion}Conclusion}
In this work, we have studied an array of coupled superconducting islands in the regime of strong charging energy. We studied the onset of topological states in this system when the Josephson energy has sublattice symmetry. This system shows mid-gap soliton states that are localized to a dimer with a weak link between the two sites using only electrostatic polarization. The topological nature of these soliton states is established quantitatively by demonstrating a non-zero winding number in the topological regime. We propose that signatures of these states are visible in the on-site AC admittance at frequencies corresponding to the energy difference from the bottom of the valence band to the mid-gap bound states. The site modulation of this spectrum coincides with that of the soliton wavefunctions. Finally, we discuss an adiabatic protocol for shuttling this bound state through a ring geometry.

\begin{acknowledgments}
 The authors wish to thank L.~F.~Banszerus, C.~M.~Marcus, and S.~Vaitiek\.{e}nas for discussion of experimental feasibility. We further acknowledge L.~Bascones, M. Burrello, O.~Mansikkam\"{a}ki, and E.~Prada for helpful discussions. We acknowledge support from the Knut and Alice Wallenberg Foundation KAW 2019.0068, European Research Council under the European Union Seventh Framework ERS-2018-SYG 810451 HERO, Spanish CM ``Talento Program" (project No. 2022-T1/IND-24070), and the University of Connecticut.
\end{acknowledgments}

\appendix
\section{\label{App:Protocol2}An illustration of a protocol P2}
    \begin{figure*}[!ht]
        \centering
        \includegraphics[scale=0.9]{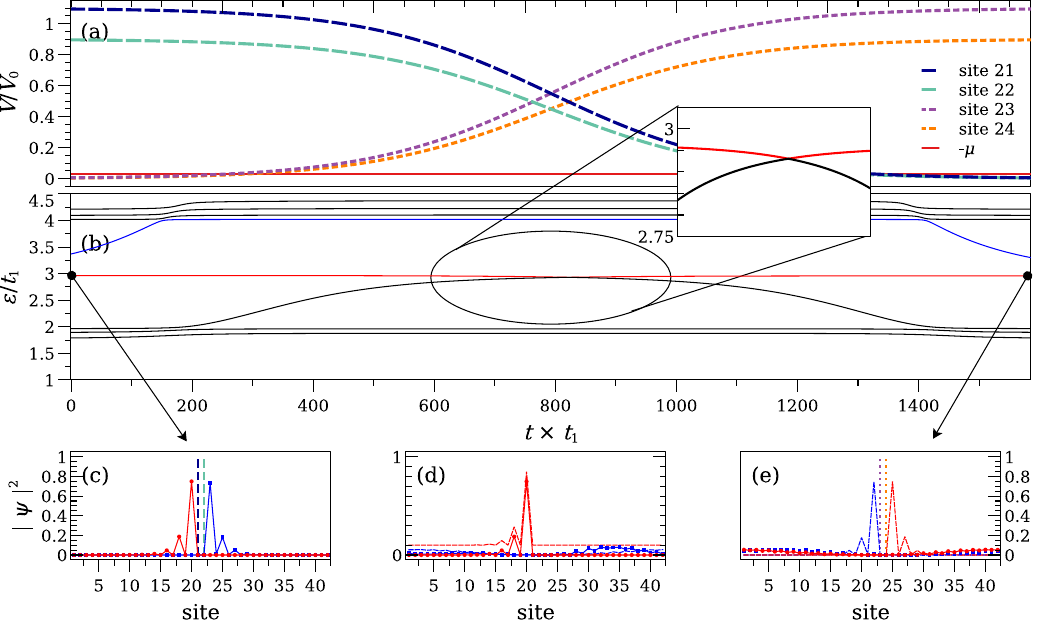}
        \caption{\label{fig:Protocol2}The protocol P2 that does not work analogous to Fig.~\ref{fig:Protocol1}, with the only difference in the relative time shift of the ramp-down and ramp-up .}
    \end{figure*}
Figure~\ref{fig:Protocol2} illustrates the protocol P2 that does not achieve an effective shuttling of the soliton dimer.
\begin{figure}
        \centering
        \includegraphics[scale=0.9]{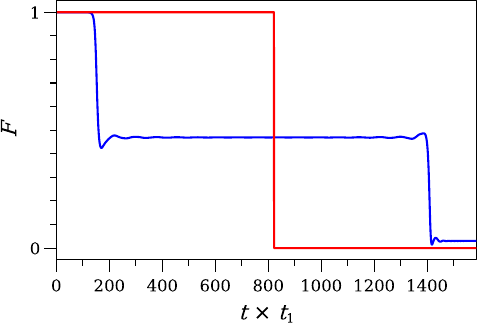}
        \caption{\label{fig:FidelityP2}The fidelity characteristics as in Fig.~\ref{fig:FidelityP1} for the protocol depicted in Fig.~\ref{fig:Protocol2}.}
    \end{figure}
The fidelity measure for P2, depicted in Fig.~\ref{fig:FidelityP2}(b), clearly shows the inability of this protocol to shuttle the state to the neighboring control dimer. A crucial difference with protocol P1 is that the blue soliton state raises and gets pinned to the bottom of the conduction band as soon as the potential on dimer 23-24 is turned on. In effect, the fidelity of the blue state drops to less than half, because of the significant overlap with only the mid-gap red state. In the middle of protocol P2, the mid-gap state exchanges role with another completely orthogonal state, which evolves from the top of the valence band, with which the former mid-gap state has zero overlap. Thus, the fidelity of the red state abruptly drops to zero. When the potential on dimer 21-22 is turned off, the higher energy state drops from the conduction band in the middle of the gap. However, it has a negligible overlap with the newly shifted mid-gap bound state, and the fidelity of the blue state has another significant drop.

\section{\label{App:Animation}Animations of shuttling protocols over the whole ring}
    \begin{video}[!h]
        \href{https://1drv.ms/i/s!AgH1CyvVitd5qwl7ZlbxD4teWj47?e=qnMcuF}{\includegraphics[width=0.9\columnwidth]{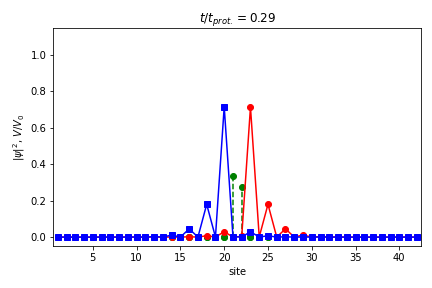}}
        \setfloatlink{https://1drv.ms/i/s!AgH1CyvVitd5qwl7ZlbxD4teWj47?e=qnMcuF}
        \caption{\label{vid:Protocol1}Animation showing the full time evolution of the soliton states according to protocol P1. Red (blue) curve are the wavefunctions initiated in the corresponding mid-gap soliton state in Fig.~\ref{fig:Protocol1}. The green dashed curved is the scaled value of the on-site control potential. The time $t_{\mathrm{prot}}$ is the time it takes for one dimer shift during protocol P1, $t_{\mathrm{prot}} \, t_{1} = 3.170 \times 10^{3}$.}
    \end{video}
The whole procedure over repeated over the ring is depicted in Video~\ref{vid:Protocol1} for P1 and Video~\ref{vid:Protocol2} for P2, respectively. We note the importance of a buffer phase of strong constant potential on the respective control dimer between the ramps, which is chosen to be of equal duration as the ramp up (down) phase.
the
    \begin{video}[!h]
        \href{https://1drv.ms/i/s!AgH1CyvVitd5qwuGapNjVZsWA_Sv?e=J7oRvn}    
        {\includegraphics[width=0.9\columnwidth]{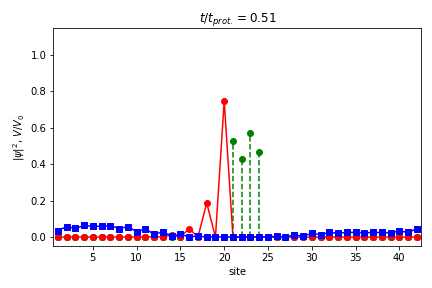}}
        \setfloatlink{https://1drv.ms/i/s!AgH1CyvVitd5qwuGapNjVZsWA_Sv?e=J7oRvn}
        \caption{\label{vid:Protocol2}Animation showing the full time evolution of the soliton states according to protocol P2. Because of the lack of a relative time shift between the ramp-down and ramp-up phases, the time it takes for one dimer shift during protocol P2 is shorter, $t_{\mathrm{prot}} \times t_{1} = 1.585 \times 10^{3}$.}
    \end{video}
    \begin{figure}[!ht]
        \centering
        \includegraphics[scale=0.9]{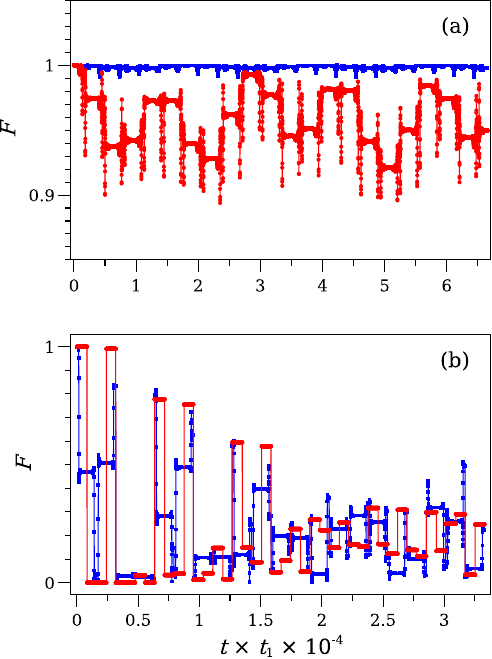}
        \caption{\label{fig:FullShuttleFidelity}The fidelity characteristics as in Figs.~\ref{fig:FidelityP1}, \ref{fig:FidelityP2} for (a) protocol P1, (b) protocol P2. The color coding refers to the corresponding mid-gap states in Figs.~\ref{fig:Protocol1}, \ref{fig:Protocol2}, respectively. Similar to Fig.~\ref{fig:FidelityP1}, note the small range of values on the vertical axis in panel (a).}
    \end{figure}
We depict the fidelity of both protocols across the whole ring in Fig.~\ref{fig:FullShuttleFidelity}. As can be seen, the fidelity is preserved for both states to appreciable levels during the whole protocol P1, with the lower energy red state having a marginally smaller value for the same reasons as in Fig.~\ref{fig:FidelityP1}. For protocol P2, on the other hand, after an accidental reemergence of fidelity during the first two steps, there is a systematic, significant drop, illustrating again the inapplicability of this protocol for shuttling the soliton states.


\begin{thebibliography}{44}%
\makeatletter
\providecommand \@ifxundefined [1]{%
 \@ifx{#1\undefined}
}%
\providecommand \@ifnum [1]{%
 \ifnum #1\expandafter \@firstoftwo
 \else \expandafter \@secondoftwo
 \fi
}%
\providecommand \@ifx [1]{%
 \ifx #1\expandafter \@firstoftwo
 \else \expandafter \@secondoftwo
 \fi
}%
\providecommand \natexlab [1]{#1}%
\providecommand \enquote  [1]{``#1''}%
\providecommand \bibnamefont  [1]{#1}%
\providecommand \bibfnamefont [1]{#1}%
\providecommand \citenamefont [1]{#1}%
\providecommand \href@noop [0]{\@secondoftwo}%
\providecommand \href [0]{\begingroup \@sanitize@url \@href}%
\providecommand \@href[1]{\@@startlink{#1}\@@href}%
\providecommand \@@href[1]{\endgroup#1\@@endlink}%
\providecommand \@sanitize@url [0]{\catcode `\\12\catcode `\$12\catcode
  `\&12\catcode `\#12\catcode `\^12\catcode `\_12\catcode `\%12\relax}%
\providecommand \@@startlink[1]{}%
\providecommand \@@endlink[0]{}%
\providecommand \url  [0]{\begingroup\@sanitize@url \@url }%
\providecommand \@url [1]{\endgroup\@href {#1}{\urlprefix }}%
\providecommand \urlprefix  [0]{URL }%
\providecommand \Eprint [0]{\href }%
\providecommand \doibase [0]{https://doi.org/}%
\providecommand \selectlanguage [0]{\@gobble}%
\providecommand \bibinfo  [0]{\@secondoftwo}%
\providecommand \bibfield  [0]{\@secondoftwo}%
\providecommand \translation [1]{[#1]}%
\providecommand \BibitemOpen [0]{}%
\providecommand \bibitemStop [0]{}%
\providecommand \bibitemNoStop [0]{.\EOS\space}%
\providecommand \EOS [0]{\spacefactor3000\relax}%
\providecommand \BibitemShut  [1]{\csname bibitem#1\endcsname}%
\let\auto@bib@innerbib\@empty
\bibitem [{\citenamefont {Neill}\ \emph {et~al.}(2021)\citenamefont {Neill},
  \citenamefont {McCourt}, \citenamefont {Mi}, \citenamefont {Jiang},
  \citenamefont {Niu}, \citenamefont {Mruczkiewicz}, \citenamefont {Aleiner},
  \citenamefont {Arute}, \citenamefont {Arya}, \citenamefont {Atalaya},
  \citenamefont {Babbush}, \citenamefont {Bardin}, \citenamefont {Barends},
  \citenamefont {Bengtsson}, \citenamefont {Bourassa}, \citenamefont
  {Broughton}, \citenamefont {Buckley}, \citenamefont {Buell}, \citenamefont
  {Burkett}, \citenamefont {Bushnell}, \citenamefont {Campero}, \citenamefont
  {Chen}, \citenamefont {Chiaro}, \citenamefont {Collins}, \citenamefont
  {Courtney}, \citenamefont {Demura}, \citenamefont {Derk}, \citenamefont
  {Dunsworth}, \citenamefont {Eppens}, \citenamefont {Erickson}, \citenamefont
  {Farhi}, \citenamefont {Fowler}, \citenamefont {Foxen}, \citenamefont
  {Gidney}, \citenamefont {Giustina}, \citenamefont {Gross}, \citenamefont
  {Harrigan}, \citenamefont {Harrington}, \citenamefont {Hilton}, \citenamefont
  {Ho}, \citenamefont {Hong}, \citenamefont {Huang}, \citenamefont {Huggins},
  \citenamefont {Isakov}, \citenamefont {Jacob-Mitos}, \citenamefont {Jeffrey},
  \citenamefont {Jones}, \citenamefont {Kafri}, \citenamefont {Kechedzhi},
  \citenamefont {Kelly}, \citenamefont {Kim}, \citenamefont {Klimov},
  \citenamefont {Korotkov}, \citenamefont {Kostritsa}, \citenamefont
  {Landhuis}, \citenamefont {Laptev}, \citenamefont {Lucero}, \citenamefont
  {Martin}, \citenamefont {McClean}, \citenamefont {McEwen}, \citenamefont
  {Megrant}, \citenamefont {Miao}, \citenamefont {Mohseni}, \citenamefont
  {Mutus}, \citenamefont {Naaman}, \citenamefont {Neeley}, \citenamefont
  {Newman}, \citenamefont {O'Brien}, \citenamefont {Opremcak}, \citenamefont
  {Ostby}, \citenamefont {Pat{\'o}}, \citenamefont {Petukhov}, \citenamefont
  {Quintana}, \citenamefont {Redd}, \citenamefont {Rubin}, \citenamefont
  {Sank}, \citenamefont {Satzinger}, \citenamefont {Shvarts}, \citenamefont
  {Strain}, \citenamefont {Szalay}, \citenamefont {Trevithick}, \citenamefont
  {Villalonga}, \citenamefont {White}, \citenamefont {Yao}, \citenamefont
  {Yeh}, \citenamefont {Zalcman}, \citenamefont {Neven}, \citenamefont {Boixo},
  \citenamefont {Ioffe}, \citenamefont {Roushan}, \citenamefont {Chen},\ and\
  \citenamefont {Smelyanskiy}}]{NeillNat2021}%
  \BibitemOpen
  \bibfield  {author} {\bibinfo {author} {\bibfnamefont {C.}~\bibnamefont
  {Neill}}, \bibinfo {author} {\bibfnamefont {T.}~\bibnamefont {McCourt}},
  \bibinfo {author} {\bibfnamefont {X.}~\bibnamefont {Mi}}, \bibinfo {author}
  {\bibfnamefont {Z.}~\bibnamefont {Jiang}}, \bibinfo {author} {\bibfnamefont
  {M.~Y.}\ \bibnamefont {Niu}}, \bibinfo {author} {\bibfnamefont
  {W.}~\bibnamefont {Mruczkiewicz}}, \bibinfo {author} {\bibfnamefont
  {I.}~\bibnamefont {Aleiner}}, \bibinfo {author} {\bibfnamefont
  {F.}~\bibnamefont {Arute}}, \bibinfo {author} {\bibfnamefont
  {K.}~\bibnamefont {Arya}}, \bibinfo {author} {\bibfnamefont {J.}~\bibnamefont
  {Atalaya}}, \bibinfo {author} {\bibfnamefont {R.}~\bibnamefont {Babbush}},
  \bibinfo {author} {\bibfnamefont {J.~C.}\ \bibnamefont {Bardin}}, \bibinfo
  {author} {\bibfnamefont {R.}~\bibnamefont {Barends}}, \bibinfo {author}
  {\bibfnamefont {A.}~\bibnamefont {Bengtsson}}, \bibinfo {author}
  {\bibfnamefont {A.}~\bibnamefont {Bourassa}}, \bibinfo {author}
  {\bibfnamefont {M.}~\bibnamefont {Broughton}}, \bibinfo {author}
  {\bibfnamefont {B.~B.}\ \bibnamefont {Buckley}}, \bibinfo {author}
  {\bibfnamefont {D.~A.}\ \bibnamefont {Buell}}, \bibinfo {author}
  {\bibfnamefont {B.}~\bibnamefont {Burkett}}, \bibinfo {author} {\bibfnamefont
  {N.}~\bibnamefont {Bushnell}}, \bibinfo {author} {\bibfnamefont
  {J.}~\bibnamefont {Campero}}, \bibinfo {author} {\bibfnamefont
  {Z.}~\bibnamefont {Chen}}, \bibinfo {author} {\bibfnamefont {B.}~\bibnamefont
  {Chiaro}}, \bibinfo {author} {\bibfnamefont {R.}~\bibnamefont {Collins}},
  \bibinfo {author} {\bibfnamefont {W.}~\bibnamefont {Courtney}}, \bibinfo
  {author} {\bibfnamefont {S.}~\bibnamefont {Demura}}, \bibinfo {author}
  {\bibfnamefont {A.~R.}\ \bibnamefont {Derk}}, \bibinfo {author}
  {\bibfnamefont {A.}~\bibnamefont {Dunsworth}}, \bibinfo {author}
  {\bibfnamefont {D.}~\bibnamefont {Eppens}}, \bibinfo {author} {\bibfnamefont
  {C.}~\bibnamefont {Erickson}}, \bibinfo {author} {\bibfnamefont
  {E.}~\bibnamefont {Farhi}}, \bibinfo {author} {\bibfnamefont {A.~G.}\
  \bibnamefont {Fowler}}, \bibinfo {author} {\bibfnamefont {B.}~\bibnamefont
  {Foxen}}, \bibinfo {author} {\bibfnamefont {C.}~\bibnamefont {Gidney}},
  \bibinfo {author} {\bibfnamefont {M.}~\bibnamefont {Giustina}}, \bibinfo
  {author} {\bibfnamefont {J.~A.}\ \bibnamefont {Gross}}, \bibinfo {author}
  {\bibfnamefont {M.~P.}\ \bibnamefont {Harrigan}}, \bibinfo {author}
  {\bibfnamefont {S.~D.}\ \bibnamefont {Harrington}}, \bibinfo {author}
  {\bibfnamefont {J.}~\bibnamefont {Hilton}}, \bibinfo {author} {\bibfnamefont
  {A.}~\bibnamefont {Ho}}, \bibinfo {author} {\bibfnamefont {S.}~\bibnamefont
  {Hong}}, \bibinfo {author} {\bibfnamefont {T.}~\bibnamefont {Huang}},
  \bibinfo {author} {\bibfnamefont {W.~J.}\ \bibnamefont {Huggins}}, \bibinfo
  {author} {\bibfnamefont {S.~V.}\ \bibnamefont {Isakov}}, \bibinfo {author}
  {\bibfnamefont {M.}~\bibnamefont {Jacob-Mitos}}, \bibinfo {author}
  {\bibfnamefont {E.}~\bibnamefont {Jeffrey}}, \bibinfo {author} {\bibfnamefont
  {C.}~\bibnamefont {Jones}}, \bibinfo {author} {\bibfnamefont
  {D.}~\bibnamefont {Kafri}}, \bibinfo {author} {\bibfnamefont
  {K.}~\bibnamefont {Kechedzhi}}, \bibinfo {author} {\bibfnamefont
  {J.}~\bibnamefont {Kelly}}, \bibinfo {author} {\bibfnamefont
  {S.}~\bibnamefont {Kim}}, \bibinfo {author} {\bibfnamefont {P.~V.}\
  \bibnamefont {Klimov}}, \bibinfo {author} {\bibfnamefont {A.~N.}\
  \bibnamefont {Korotkov}}, \bibinfo {author} {\bibfnamefont {F.}~\bibnamefont
  {Kostritsa}}, \bibinfo {author} {\bibfnamefont {D.}~\bibnamefont {Landhuis}},
  \bibinfo {author} {\bibfnamefont {P.}~\bibnamefont {Laptev}}, \bibinfo
  {author} {\bibfnamefont {E.}~\bibnamefont {Lucero}}, \bibinfo {author}
  {\bibfnamefont {O.}~\bibnamefont {Martin}}, \bibinfo {author} {\bibfnamefont
  {J.~R.}\ \bibnamefont {McClean}}, \bibinfo {author} {\bibfnamefont
  {M.}~\bibnamefont {McEwen}}, \bibinfo {author} {\bibfnamefont
  {A.}~\bibnamefont {Megrant}}, \bibinfo {author} {\bibfnamefont {K.~C.}\
  \bibnamefont {Miao}}, \bibinfo {author} {\bibfnamefont {M.}~\bibnamefont
  {Mohseni}}, \bibinfo {author} {\bibfnamefont {J.}~\bibnamefont {Mutus}},
  \bibinfo {author} {\bibfnamefont {O.}~\bibnamefont {Naaman}}, \bibinfo
  {author} {\bibfnamefont {M.}~\bibnamefont {Neeley}}, \bibinfo {author}
  {\bibfnamefont {M.}~\bibnamefont {Newman}}, \bibinfo {author} {\bibfnamefont
  {T.~E.}\ \bibnamefont {O'Brien}}, \bibinfo {author} {\bibfnamefont
  {A.}~\bibnamefont {Opremcak}}, \bibinfo {author} {\bibfnamefont
  {E.}~\bibnamefont {Ostby}}, \bibinfo {author} {\bibfnamefont
  {B.}~\bibnamefont {Pat{\'o}}}, \bibinfo {author} {\bibfnamefont
  {A.}~\bibnamefont {Petukhov}}, \bibinfo {author} {\bibfnamefont
  {C.}~\bibnamefont {Quintana}}, \bibinfo {author} {\bibfnamefont
  {N.}~\bibnamefont {Redd}}, \bibinfo {author} {\bibfnamefont {N.~C.}\
  \bibnamefont {Rubin}}, \bibinfo {author} {\bibfnamefont {D.}~\bibnamefont
  {Sank}}, \bibinfo {author} {\bibfnamefont {K.~J.}\ \bibnamefont {Satzinger}},
  \bibinfo {author} {\bibfnamefont {V.}~\bibnamefont {Shvarts}}, \bibinfo
  {author} {\bibfnamefont {D.}~\bibnamefont {Strain}}, \bibinfo {author}
  {\bibfnamefont {M.}~\bibnamefont {Szalay}}, \bibinfo {author} {\bibfnamefont
  {M.~D.}\ \bibnamefont {Trevithick}}, \bibinfo {author} {\bibfnamefont
  {B.}~\bibnamefont {Villalonga}}, \bibinfo {author} {\bibfnamefont {T.~C.}\
  \bibnamefont {White}}, \bibinfo {author} {\bibfnamefont {Z.}~\bibnamefont
  {Yao}}, \bibinfo {author} {\bibfnamefont {P.}~\bibnamefont {Yeh}}, \bibinfo
  {author} {\bibfnamefont {A.}~\bibnamefont {Zalcman}}, \bibinfo {author}
  {\bibfnamefont {H.}~\bibnamefont {Neven}}, \bibinfo {author} {\bibfnamefont
  {S.}~\bibnamefont {Boixo}}, \bibinfo {author} {\bibfnamefont {L.~B.}\
  \bibnamefont {Ioffe}}, \bibinfo {author} {\bibfnamefont {P.}~\bibnamefont
  {Roushan}}, \bibinfo {author} {\bibfnamefont {Y.}~\bibnamefont {Chen}},\ and\
  \bibinfo {author} {\bibfnamefont {V.}~\bibnamefont {Smelyanskiy}},\
  }\bibfield  {title} {\bibinfo {title} {Accurately computing the electronic
  properties of a quantum ring},\ }\href
  {https://doi.org/10.1038/s41586-021-03576-2} {\bibfield  {journal} {\bibinfo
  {journal} {Nature}\ }\textbf {\bibinfo {volume} {594}},\ \bibinfo {pages}
  {508} (\bibinfo {year} {2021})}\BibitemShut {NoStop}%
\bibitem [{\citenamefont {Georgescu}\ \emph {et~al.}(2014)\citenamefont
  {Georgescu}, \citenamefont {Ashhab},\ and\ \citenamefont
  {Nori}}]{GeorgescuRMP2014}%
  \BibitemOpen
  \bibfield  {author} {\bibinfo {author} {\bibfnamefont {I.~M.}\ \bibnamefont
  {Georgescu}}, \bibinfo {author} {\bibfnamefont {S.}~\bibnamefont {Ashhab}},\
  and\ \bibinfo {author} {\bibfnamefont {F.}~\bibnamefont {Nori}},\ }\bibfield
  {title} {\bibinfo {title} {{Quantum simulation}},\ }\href
  {https://doi.org/10.1103/RevModPhys.86.153} {\bibfield  {journal} {\bibinfo
  {journal} {Reviews of Modern Physics}\ }\textbf {\bibinfo {volume} {86}},\
  \bibinfo {pages} {153} (\bibinfo {year} {2014})}\BibitemShut {NoStop}%
\bibitem [{\citenamefont {Monroe}\ \emph {et~al.}(2021)\citenamefont {Monroe},
  \citenamefont {Campbell}, \citenamefont {Duan}, \citenamefont {Gong},
  \citenamefont {Gorshkov}, \citenamefont {Hess}, \citenamefont {Islam},
  \citenamefont {Kim}, \citenamefont {Linke}, \citenamefont {Pagano},
  \citenamefont {Richerme}, \citenamefont {Senko},\ and\ \citenamefont
  {Yao}}]{MonroeRMP2021}%
  \BibitemOpen
  \bibfield  {author} {\bibinfo {author} {\bibfnamefont {C.}~\bibnamefont
  {Monroe}}, \bibinfo {author} {\bibfnamefont {W.~C.}\ \bibnamefont
  {Campbell}}, \bibinfo {author} {\bibfnamefont {L.-M.}\ \bibnamefont {Duan}},
  \bibinfo {author} {\bibfnamefont {Z.-X.}\ \bibnamefont {Gong}}, \bibinfo
  {author} {\bibfnamefont {A.~V.}\ \bibnamefont {Gorshkov}}, \bibinfo {author}
  {\bibfnamefont {P.~W.}\ \bibnamefont {Hess}}, \bibinfo {author}
  {\bibfnamefont {R.}~\bibnamefont {Islam}}, \bibinfo {author} {\bibfnamefont
  {K.}~\bibnamefont {Kim}}, \bibinfo {author} {\bibfnamefont {N.~M.}\
  \bibnamefont {Linke}}, \bibinfo {author} {\bibfnamefont {G.}~\bibnamefont
  {Pagano}}, \bibinfo {author} {\bibfnamefont {P.}~\bibnamefont {Richerme}},
  \bibinfo {author} {\bibfnamefont {C.}~\bibnamefont {Senko}},\ and\ \bibinfo
  {author} {\bibfnamefont {N.~Y.}\ \bibnamefont {Yao}},\ }\bibfield  {title}
  {\bibinfo {title} {{Programmable quantum simulations of spin systems with
  trapped ions}},\ }\href {https://doi.org/10.1103/RevModPhys.93.025001}
  {\bibfield  {journal} {\bibinfo  {journal} {Reviews of Modern Physics}\
  }\textbf {\bibinfo {volume} {93}},\ \bibinfo {pages} {025001} (\bibinfo
  {year} {2021})}\BibitemShut {NoStop}%
\bibitem [{\citenamefont {Bloch}\ \emph {et~al.}(2012)\citenamefont {Bloch},
  \citenamefont {Dalibard},\ and\ \citenamefont {Nascimb{\`e}ne}}]{Bloch2012}%
  \BibitemOpen
  \bibfield  {author} {\bibinfo {author} {\bibfnamefont {I.}~\bibnamefont
  {Bloch}}, \bibinfo {author} {\bibfnamefont {J.}~\bibnamefont {Dalibard}},\
  and\ \bibinfo {author} {\bibfnamefont {S.}~\bibnamefont {Nascimb{\`e}ne}},\
  }\bibfield  {title} {\bibinfo {title} {{Quantum simulations with ultracold
  quantum gases}},\ }\href {https://doi.org/10.1038/nphys2259} {\bibfield
  {journal} {\bibinfo  {journal} {Nature Physics}\ }\textbf {\bibinfo {volume}
  {8}},\ \bibinfo {pages} {267} (\bibinfo {year} {2012})}\BibitemShut {NoStop}%
\bibitem [{\citenamefont {Schmidt}\ and\ \citenamefont
  {Koch}(2013)}]{Schmidt2013}%
  \BibitemOpen
  \bibfield  {author} {\bibinfo {author} {\bibfnamefont {S.}~\bibnamefont
  {Schmidt}}\ and\ \bibinfo {author} {\bibfnamefont {J.}~\bibnamefont {Koch}},\
  }\bibfield  {title} {\bibinfo {title} {{Circuit QED lattices: Towards quantum
  simulation with superconducting circuits}},\ }\href
  {https://doi.org/10.1002/andp.201200261} {\bibfield  {journal} {\bibinfo
  {journal} {Annalen der Physik}\ }\textbf {\bibinfo {volume} {525}},\ \bibinfo
  {pages} {395} (\bibinfo {year} {2013})}\BibitemShut {NoStop}%
\bibitem [{\citenamefont {Geerligs}\ \emph {et~al.}(1989)\citenamefont
  {Geerligs}, \citenamefont {Peters}, \citenamefont {de~Groot}, \citenamefont
  {Verbruggen},\ and\ \citenamefont {Mooij}}]{GeerligsPRL1989}%
  \BibitemOpen
  \bibfield  {author} {\bibinfo {author} {\bibfnamefont {L.~J.}\ \bibnamefont
  {Geerligs}}, \bibinfo {author} {\bibfnamefont {M.}~\bibnamefont {Peters}},
  \bibinfo {author} {\bibfnamefont {L.~E.~M.}\ \bibnamefont {de~Groot}},
  \bibinfo {author} {\bibfnamefont {A.}~\bibnamefont {Verbruggen}},\ and\
  \bibinfo {author} {\bibfnamefont {J.~E.}\ \bibnamefont {Mooij}},\ }\bibfield
  {title} {\bibinfo {title} {{Charging effects and quantum coherence in regular
  Josephson junction arrays}},\ }\href
  {https://doi.org/10.1103/PhysRevLett.63.326} {\bibfield  {journal} {\bibinfo
  {journal} {Physical Review Letters}\ }\textbf {\bibinfo {volume} {63}},\
  \bibinfo {pages} {326} (\bibinfo {year} {1989})}\BibitemShut {NoStop}%
\bibitem [{\citenamefont {Chen}\ \emph {et~al.}(1995)\citenamefont {Chen},
  \citenamefont {Delsing}, \citenamefont {Haviland}, \citenamefont {Harada},\
  and\ \citenamefont {Claeson}}]{ChenPRB1995}%
  \BibitemOpen
  \bibfield  {author} {\bibinfo {author} {\bibfnamefont {C.~D.}\ \bibnamefont
  {Chen}}, \bibinfo {author} {\bibfnamefont {P.}~\bibnamefont {Delsing}},
  \bibinfo {author} {\bibfnamefont {D.~B.}\ \bibnamefont {Haviland}}, \bibinfo
  {author} {\bibfnamefont {Y.}~\bibnamefont {Harada}},\ and\ \bibinfo {author}
  {\bibfnamefont {T.}~\bibnamefont {Claeson}},\ }\bibfield  {title} {\bibinfo
  {title} {{Scaling behavior of the magnetic-field-tuned
  superconductor-insulator transition in two-dimensional Josephson-junction
  arrays}},\ }\href {https://doi.org/10.1103/PhysRevB.51.15645} {\bibfield
  {journal} {\bibinfo  {journal} {Physical Review B}\ }\textbf {\bibinfo
  {volume} {51}},\ \bibinfo {pages} {15645(R)} (\bibinfo {year}
  {1995})}\BibitemShut {NoStop}%
\bibitem [{\citenamefont {van~der Zant}\ \emph {et~al.}(1996)\citenamefont
  {van~der Zant}, \citenamefont {Elion}, \citenamefont {Geerligs},\ and\
  \citenamefont {Mooij}}]{ZantPRB1996}%
  \BibitemOpen
  \bibfield  {author} {\bibinfo {author} {\bibfnamefont {H.~S.~J.}\
  \bibnamefont {van~der Zant}}, \bibinfo {author} {\bibfnamefont {W.~J.}\
  \bibnamefont {Elion}}, \bibinfo {author} {\bibfnamefont {L.~J.}\ \bibnamefont
  {Geerligs}},\ and\ \bibinfo {author} {\bibfnamefont {J.~E.}\ \bibnamefont
  {Mooij}},\ }\bibfield  {title} {\bibinfo {title} {{Quantum phase transitions
  in two dimensions: Experiments in Josephson-junction arrays}},\ }\href
  {https://doi.org/10.1103/PhysRevB.54.10081} {\bibfield  {journal} {\bibinfo
  {journal} {Physical Review B}\ }\textbf {\bibinfo {volume} {54}},\ \bibinfo
  {pages} {10081} (\bibinfo {year} {1996})}\BibitemShut {NoStop}%
\bibitem [{\citenamefont {Basko}\ \emph {et~al.}(2020)\citenamefont {Basko},
  \citenamefont {Pfeiffer}, \citenamefont {Adamus}, \citenamefont {Holzmann},\
  and\ \citenamefont {Hekking}}]{BaskoPRB2020}%
  \BibitemOpen
  \bibfield  {author} {\bibinfo {author} {\bibfnamefont {D.~M.}\ \bibnamefont
  {Basko}}, \bibinfo {author} {\bibfnamefont {F.}~\bibnamefont {Pfeiffer}},
  \bibinfo {author} {\bibfnamefont {P.}~\bibnamefont {Adamus}}, \bibinfo
  {author} {\bibfnamefont {M.}~\bibnamefont {Holzmann}},\ and\ \bibinfo
  {author} {\bibfnamefont {F.~W.~J.}\ \bibnamefont {Hekking}},\ }\bibfield
  {title} {\bibinfo {title} {{Superconductor-insulator transition in Josephson
  junction chains by quantum Monte Carlo calculations}},\ }\href
  {https://doi.org/10.1103/PhysRevB.101.024518} {\bibfield  {journal} {\bibinfo
   {journal} {Physical Review B}\ }\textbf {\bibinfo {volume} {101}},\ \bibinfo
  {pages} {024518} (\bibinfo {year} {2020})}\BibitemShut {NoStop}%
\bibitem [{\citenamefont {B{\o}ttcher}\ \emph {et~al.}(2018)\citenamefont
  {B{\o}ttcher}, \citenamefont {Nichele}, \citenamefont {Kjaergaard},
  \citenamefont {Suominen}, \citenamefont {Shabani}, \citenamefont
  {Palmstr{\o}m},\ and\ \citenamefont {Marcus}}]{BoettcherNat2018}%
  \BibitemOpen
  \bibfield  {author} {\bibinfo {author} {\bibfnamefont {C.~G.~L.}\
  \bibnamefont {B{\o}ttcher}}, \bibinfo {author} {\bibfnamefont
  {F.}~\bibnamefont {Nichele}}, \bibinfo {author} {\bibfnamefont
  {M.}~\bibnamefont {Kjaergaard}}, \bibinfo {author} {\bibfnamefont {H.~J.}\
  \bibnamefont {Suominen}}, \bibinfo {author} {\bibfnamefont {J.}~\bibnamefont
  {Shabani}}, \bibinfo {author} {\bibfnamefont {C.~J.}\ \bibnamefont
  {Palmstr{\o}m}},\ and\ \bibinfo {author} {\bibfnamefont {C.~M.}\ \bibnamefont
  {Marcus}},\ }\bibfield  {title} {\bibinfo {title} {{Superconducting,
  insulating and anomalous metallic regimes in a gated two-dimensional
  semiconductor--superconductor array}},\ }\href
  {https://doi.org/10.1038/s41567-018-0259-9} {\bibfield  {journal} {\bibinfo
  {journal} {Nature Physics}\ }\textbf {\bibinfo {volume} {14}},\ \bibinfo
  {pages} {1138} (\bibinfo {year} {2018})}\BibitemShut {NoStop}%
\bibitem [{\citenamefont {Rakhmanov}\ \emph {et~al.}(2008)\citenamefont
  {Rakhmanov}, \citenamefont {Zagoskin}, \citenamefont {Savel'ev},\ and\
  \citenamefont {Nori}}]{RakhmanovPRB2008}%
  \BibitemOpen
  \bibfield  {author} {\bibinfo {author} {\bibfnamefont {A.~L.}\ \bibnamefont
  {Rakhmanov}}, \bibinfo {author} {\bibfnamefont {A.~M.}\ \bibnamefont
  {Zagoskin}}, \bibinfo {author} {\bibfnamefont {S.}~\bibnamefont {Savel'ev}},\
  and\ \bibinfo {author} {\bibfnamefont {F.}~\bibnamefont {Nori}},\ }\bibfield
  {title} {\bibinfo {title} {{Quantum metamaterials: Electromagnetic waves in a
  Josephson qubit line}},\ }\href {https://doi.org/10.1103/PhysRevB.77.144507}
  {\bibfield  {journal} {\bibinfo  {journal} {Physical Review B}\ }\textbf
  {\bibinfo {volume} {77}},\ \bibinfo {pages} {144507} (\bibinfo {year}
  {2008})}\BibitemShut {NoStop}%
\bibitem [{\citenamefont {Macha}\ \emph {et~al.}(2014)\citenamefont {Macha},
  \citenamefont {Oelsner}, \citenamefont {Reiner}, \citenamefont {Marthaler},
  \citenamefont {Andr{\'e}}, \citenamefont {Sch{\"o}n}, \citenamefont
  {H{\"u}bner}, \citenamefont {Meyer}, \citenamefont {Il'ichev},\ and\
  \citenamefont {Ustinov}}]{MachaNatComm2014}%
  \BibitemOpen
  \bibfield  {author} {\bibinfo {author} {\bibfnamefont {P.}~\bibnamefont
  {Macha}}, \bibinfo {author} {\bibfnamefont {G.}~\bibnamefont {Oelsner}},
  \bibinfo {author} {\bibfnamefont {J.-M.}\ \bibnamefont {Reiner}}, \bibinfo
  {author} {\bibfnamefont {M.}~\bibnamefont {Marthaler}}, \bibinfo {author}
  {\bibfnamefont {S.}~\bibnamefont {Andr{\'e}}}, \bibinfo {author}
  {\bibfnamefont {G.}~\bibnamefont {Sch{\"o}n}}, \bibinfo {author}
  {\bibfnamefont {U.}~\bibnamefont {H{\"u}bner}}, \bibinfo {author}
  {\bibfnamefont {H.-G.}\ \bibnamefont {Meyer}}, \bibinfo {author}
  {\bibfnamefont {E.}~\bibnamefont {Il'ichev}},\ and\ \bibinfo {author}
  {\bibfnamefont {A.~V.}\ \bibnamefont {Ustinov}},\ }\bibfield  {title}
  {\bibinfo {title} {{Implementation of a quantum metamaterial using
  superconducting qubits}},\ }\href {https://doi.org/10.1038/ncomms6146}
  {\bibfield  {journal} {\bibinfo  {journal} {Nature Communications}\ }\textbf
  {\bibinfo {volume} {5}},\ \bibinfo {pages} {5146} (\bibinfo {year}
  {2014})}\BibitemShut {NoStop}%
\bibitem [{\citenamefont {Brehm}\ \emph {et~al.}(2022)\citenamefont {Brehm},
  \citenamefont {Gebauer}, \citenamefont {Stehli}, \citenamefont {Poddubny},
  \citenamefont {Sander}, \citenamefont {Rotzinger},\ and\ \citenamefont
  {Ustinov}}]{BrehmAPL2022}%
  \BibitemOpen
  \bibfield  {author} {\bibinfo {author} {\bibfnamefont {J.~D.}\ \bibnamefont
  {Brehm}}, \bibinfo {author} {\bibfnamefont {R.}~\bibnamefont {Gebauer}},
  \bibinfo {author} {\bibfnamefont {A.}~\bibnamefont {Stehli}}, \bibinfo
  {author} {\bibfnamefont {A.~N.}\ \bibnamefont {Poddubny}}, \bibinfo {author}
  {\bibfnamefont {O.}~\bibnamefont {Sander}}, \bibinfo {author} {\bibfnamefont
  {H.}~\bibnamefont {Rotzinger}},\ and\ \bibinfo {author} {\bibfnamefont
  {A.~V.}\ \bibnamefont {Ustinov}},\ }\bibfield  {title} {\bibinfo {title}
  {{Slowing down light in a qubit metamaterial}},\ }\href
  {https://doi.org/10.1063/5.0122003} {\bibfield  {journal} {\bibinfo
  {journal} {Applied Physics Letters}\ }\textbf {\bibinfo {volume} {121}},\
  \bibinfo {pages} {204001} (\bibinfo {year} {2022})}\BibitemShut {NoStop}%
\bibitem [{\citenamefont {Su}\ \emph {et~al.}(1979)\citenamefont {Su},
  \citenamefont {Schrieffer},\ and\ \citenamefont {Heeger}}]{su1979}%
  \BibitemOpen
  \bibfield  {author} {\bibinfo {author} {\bibfnamefont {W.~P.}\ \bibnamefont
  {Su}}, \bibinfo {author} {\bibfnamefont {J.~R.}\ \bibnamefont {Schrieffer}},\
  and\ \bibinfo {author} {\bibfnamefont {A.~J.}\ \bibnamefont {Heeger}},\
  }\bibfield  {title} {\bibinfo {title} {{Solitons in Polyacetylene}},\ }\href
  {https://doi.org/10.1103/PhysRevLett.42.1698} {\bibfield  {journal} {\bibinfo
   {journal} {Physical Review Letters}\ }\textbf {\bibinfo {volume} {42}},\
  \bibinfo {pages} {1698} (\bibinfo {year} {1979})}\BibitemShut {NoStop}%
\bibitem [{\citenamefont {Heeger}\ \emph {et~al.}(1988)\citenamefont {Heeger},
  \citenamefont {Kivelson}, \citenamefont {Schrieffer},\ and\ \citenamefont
  {Su}}]{heeger1988}%
  \BibitemOpen
  \bibfield  {author} {\bibinfo {author} {\bibfnamefont {A.~J.}\ \bibnamefont
  {Heeger}}, \bibinfo {author} {\bibfnamefont {S.}~\bibnamefont {Kivelson}},
  \bibinfo {author} {\bibfnamefont {J.~R.}\ \bibnamefont {Schrieffer}},\ and\
  \bibinfo {author} {\bibfnamefont {W.-P.}\ \bibnamefont {Su}},\ }\bibfield
  {title} {\bibinfo {title} {{Solitons in conducting polymers}},\ }\href
  {https://doi.org/10.1103/RevModPhys.60.781} {\bibfield  {journal} {\bibinfo
  {journal} {Reviews of Modern Physics}\ }\textbf {\bibinfo {volume} {60}},\
  \bibinfo {pages} {781} (\bibinfo {year} {1988})}\BibitemShut {NoStop}%
\bibitem [{\citenamefont {Asb\'{o}th}\ \emph {et~al.}(2016)\citenamefont
  {Asb\'{o}th}, \citenamefont {Oroszl\'{a}ny},\ and\ \citenamefont
  {P\'{a}lyi}}]{asboth2016}%
  \BibitemOpen
  \bibfield  {author} {\bibinfo {author} {\bibfnamefont {J.~K.}\ \bibnamefont
  {Asb\'{o}th}}, \bibinfo {author} {\bibfnamefont {L.}~\bibnamefont
  {Oroszl\'{a}ny}},\ and\ \bibinfo {author} {\bibfnamefont {A.}~\bibnamefont
  {P\'{a}lyi}},\ }\href {https://doi.org/10.1007/978-3-319-25607-8} {\emph
  {\bibinfo {title} {{A Short Course on Topological Insulators}}}},\ \bibinfo
  {edition} {1st}\ ed.,\ \bibinfo {series} {Lecture Notes in Physics}, Vol.\
  \bibinfo {volume} {919}\ (\bibinfo  {publisher} {Springer Cham},\ \bibinfo
  {year} {2016})\BibitemShut {NoStop}%
\bibitem [{\citenamefont {Takayama}\ \emph {et~al.}(1980)\citenamefont
  {Takayama}, \citenamefont {Lin-Liu},\ and\ \citenamefont
  {Maki}}]{takayama1980}%
  \BibitemOpen
  \bibfield  {author} {\bibinfo {author} {\bibfnamefont {H.}~\bibnamefont
  {Takayama}}, \bibinfo {author} {\bibfnamefont {Y.~R.}\ \bibnamefont
  {Lin-Liu}},\ and\ \bibinfo {author} {\bibfnamefont {K.}~\bibnamefont
  {Maki}},\ }\bibfield  {title} {\bibinfo {title} {{Continuum model for
  solitons in polyacetylene}},\ }\href
  {https://doi.org/10.1103/PhysRevB.21.2388} {\bibfield  {journal} {\bibinfo
  {journal} {Physical Review B}\ }\textbf {\bibinfo {volume} {21}},\ \bibinfo
  {pages} {2388} (\bibinfo {year} {1980})}\BibitemShut {NoStop}%
\bibitem [{\citenamefont {Chen}\ \emph {et~al.}(2021)\citenamefont {Chen},
  \citenamefont {Chen}, \citenamefont {Ren}, \citenamefont {Gong},\ and\
  \citenamefont {Guo}}]{chen2021}%
  \BibitemOpen
  \bibfield  {author} {\bibinfo {author} {\bibfnamefont {Y.}~\bibnamefont
  {Chen}}, \bibinfo {author} {\bibfnamefont {X.}~\bibnamefont {Chen}}, \bibinfo
  {author} {\bibfnamefont {X.}~\bibnamefont {Ren}}, \bibinfo {author}
  {\bibfnamefont {M.}~\bibnamefont {Gong}},\ and\ \bibinfo {author}
  {\bibfnamefont {G.-c.}\ \bibnamefont {Guo}},\ }\bibfield  {title} {\bibinfo
  {title} {{Tight-binding model in optical waveguides: Design principle and
  transferability for simulation of complex photonics networks}},\ }\href
  {https://doi.org/10.1103/PhysRevA.104.023501} {\bibfield  {journal} {\bibinfo
   {journal} {Physical Review A}\ }\textbf {\bibinfo {volume} {104}},\ \bibinfo
  {pages} {023501} (\bibinfo {year} {2021})}\BibitemShut {NoStop}%
\bibitem [{\citenamefont {Liu}\ \emph {et~al.}(2020)\citenamefont {Liu},
  \citenamefont {Chen},\ and\ \citenamefont {Xu}}]{liu2020}%
  \BibitemOpen
  \bibfield  {author} {\bibinfo {author} {\bibfnamefont {Y.}~\bibnamefont
  {Liu}}, \bibinfo {author} {\bibfnamefont {X.}~\bibnamefont {Chen}},\ and\
  \bibinfo {author} {\bibfnamefont {Y.}~\bibnamefont {Xu}},\ }\bibfield
  {title} {\bibinfo {title} {{Topological Phononics: From Fundamental Models to
  Real Materials}},\ }\href
  {https://doi.org/https://doi.org/10.1002/adfm.201904784} {\bibfield
  {journal} {\bibinfo  {journal} {Advanced Functional Materials}\ }\textbf
  {\bibinfo {volume} {30}},\ \bibinfo {pages} {1904784} (\bibinfo {year}
  {2020})}\BibitemShut {NoStop}%
\bibitem [{\citenamefont {Mei}\ \emph {et~al.}(2019)\citenamefont {Mei},
  \citenamefont {Chen}, \citenamefont {Goldman}, \citenamefont {Xiao},\ and\
  \citenamefont {Jia}}]{mei2019}%
  \BibitemOpen
  \bibfield  {author} {\bibinfo {author} {\bibfnamefont {F.}~\bibnamefont
  {Mei}}, \bibinfo {author} {\bibfnamefont {G.}~\bibnamefont {Chen}}, \bibinfo
  {author} {\bibfnamefont {N.}~\bibnamefont {Goldman}}, \bibinfo {author}
  {\bibfnamefont {L.}~\bibnamefont {Xiao}},\ and\ \bibinfo {author}
  {\bibfnamefont {S.}~\bibnamefont {Jia}},\ }\bibfield  {title} {\bibinfo
  {title} {{Topological magnon insulator and quantized pumps from
  strongly-interacting bosons in optical superlattices}},\ }\href
  {https://doi.org/10.1088/1367-2630/ab3d93} {\bibfield  {journal} {\bibinfo
  {journal} {New Journal of Physics}\ }\textbf {\bibinfo {volume} {21}},\
  \bibinfo {pages} {095002} (\bibinfo {year} {2019})}\BibitemShut {NoStop}%
\bibitem [{\citenamefont {Li}\ and\ \citenamefont {Cheng}(2021)}]{li2021}%
  \BibitemOpen
  \bibfield  {author} {\bibinfo {author} {\bibfnamefont {Y.-H.}\ \bibnamefont
  {Li}}\ and\ \bibinfo {author} {\bibfnamefont {R.}~\bibnamefont {Cheng}},\
  }\bibfield  {title} {\bibinfo {title} {{Magnonic Su-Schrieffer-Heeger model
  in honeycomb ferromagnets}},\ }\href
  {https://doi.org/10.1103/PhysRevB.103.014407} {\bibfield  {journal} {\bibinfo
   {journal} {Physical Review B}\ }\textbf {\bibinfo {volume} {103}},\ \bibinfo
  {pages} {014407} (\bibinfo {year} {2021})}\BibitemShut {NoStop}%
\bibitem [{\citenamefont {Wei}\ \emph {et~al.}(2022)\citenamefont {Wei},
  \citenamefont {Ni}, \citenamefont {Zheng}, \citenamefont {Liu},\ and\
  \citenamefont {Zou}}]{WeiJPCM2022}%
  \BibitemOpen
  \bibfield  {author} {\bibinfo {author} {\bibfnamefont {P.-T.}\ \bibnamefont
  {Wei}}, \bibinfo {author} {\bibfnamefont {J.-Y.}\ \bibnamefont {Ni}},
  \bibinfo {author} {\bibfnamefont {X.-M.}\ \bibnamefont {Zheng}}, \bibinfo
  {author} {\bibfnamefont {D.-Y.}\ \bibnamefont {Liu}},\ and\ \bibinfo {author}
  {\bibfnamefont {L.-J.}\ \bibnamefont {Zou}},\ }\bibfield  {title} {\bibinfo
  {title} {{Topological magnons in one-dimensional ferromagnetic
  Su--Schrieffer--Heeger model with anisotropic interaction}},\ }\href
  {https://doi.org/10.1088/1361-648X/ac99cb} {\bibfield  {journal} {\bibinfo
  {journal} {Journal of Physics: Condensed Matter}\ }\textbf {\bibinfo {volume}
  {34}},\ \bibinfo {pages} {495801} (\bibinfo {year} {2022})}\BibitemShut
  {NoStop}%
\bibitem [{\citenamefont {Meier}\ \emph {et~al.}(2016)\citenamefont {Meier},
  \citenamefont {An},\ and\ \citenamefont {Gadway}}]{meier2016}%
  \BibitemOpen
  \bibfield  {author} {\bibinfo {author} {\bibfnamefont {E.~J.}\ \bibnamefont
  {Meier}}, \bibinfo {author} {\bibfnamefont {F.~A.}\ \bibnamefont {An}},\ and\
  \bibinfo {author} {\bibfnamefont {B.}~\bibnamefont {Gadway}},\ }\bibfield
  {title} {\bibinfo {title} {{Observation of the topological soliton state in
  the Su--Schrieffer--Heeger model}},\ }\href
  {https://doi.org/10.1038/ncomms13986} {\bibfield  {journal} {\bibinfo
  {journal} {Nature Communications}\ }\textbf {\bibinfo {volume} {7}},\
  \bibinfo {pages} {13986} (\bibinfo {year} {2016})}\BibitemShut {NoStop}%
\bibitem [{\citenamefont {de~L\'{e}s\'{e}leuc}\ \emph
  {et~al.}(2019)\citenamefont {de~L\'{e}s\'{e}leuc}, \citenamefont {Lienhard},
  \citenamefont {Scholl}, \citenamefont {Barredo}, \citenamefont {Weber},
  \citenamefont {Lang}, \citenamefont {B\"{u}chler}, \citenamefont {Lahaye},\
  and\ \citenamefont {Browaeys}}]{deleseleuc2019}%
  \BibitemOpen
  \bibfield  {author} {\bibinfo {author} {\bibfnamefont {S.}~\bibnamefont
  {de~L\'{e}s\'{e}leuc}}, \bibinfo {author} {\bibfnamefont {V.}~\bibnamefont
  {Lienhard}}, \bibinfo {author} {\bibfnamefont {P.}~\bibnamefont {Scholl}},
  \bibinfo {author} {\bibfnamefont {D.}~\bibnamefont {Barredo}}, \bibinfo
  {author} {\bibfnamefont {S.}~\bibnamefont {Weber}}, \bibinfo {author}
  {\bibfnamefont {N.}~\bibnamefont {Lang}}, \bibinfo {author} {\bibfnamefont
  {H.~P.}\ \bibnamefont {B\"{u}chler}}, \bibinfo {author} {\bibfnamefont
  {T.}~\bibnamefont {Lahaye}},\ and\ \bibinfo {author} {\bibfnamefont
  {A.}~\bibnamefont {Browaeys}},\ }\bibfield  {title} {\bibinfo {title}
  {{Observation of a symmetry-protected topological phase of interacting bosons
  with Rydberg atoms}},\ }\href {https://doi.org/10.1126/science.aav9105}
  {\bibfield  {journal} {\bibinfo  {journal} {Science}\ }\textbf {\bibinfo
  {volume} {365}},\ \bibinfo {pages} {775} (\bibinfo {year}
  {2019})}\BibitemShut {NoStop}%
\bibitem [{\citenamefont {Atala}\ \emph {et~al.}(2013)\citenamefont {Atala},
  \citenamefont {Aidelsburger}, \citenamefont {Barreiro}, \citenamefont
  {Abanin}, \citenamefont {Kitagawa}, \citenamefont {Demler},\ and\
  \citenamefont {Bloch}}]{atala2013}%
  \BibitemOpen
  \bibfield  {author} {\bibinfo {author} {\bibfnamefont {M.}~\bibnamefont
  {Atala}}, \bibinfo {author} {\bibfnamefont {M.}~\bibnamefont {Aidelsburger}},
  \bibinfo {author} {\bibfnamefont {J.~T.}\ \bibnamefont {Barreiro}}, \bibinfo
  {author} {\bibfnamefont {D.}~\bibnamefont {Abanin}}, \bibinfo {author}
  {\bibfnamefont {T.}~\bibnamefont {Kitagawa}}, \bibinfo {author}
  {\bibfnamefont {E.}~\bibnamefont {Demler}},\ and\ \bibinfo {author}
  {\bibfnamefont {I.}~\bibnamefont {Bloch}},\ }\bibfield  {title} {\bibinfo
  {title} {{Direct measurement of the Zak phase in topological Bloch bands}},\
  }\href {https://doi.org/10.1038/nphys2790} {\bibfield  {journal} {\bibinfo
  {journal} {Nature Physics}\ }\textbf {\bibinfo {volume} {9}},\ \bibinfo
  {pages} {795} (\bibinfo {year} {2013})}\BibitemShut {NoStop}%
\bibitem [{\citenamefont {Lohse}\ \emph {et~al.}(2016)\citenamefont {Lohse},
  \citenamefont {Schweizer}, \citenamefont {Zilberberg}, \citenamefont
  {Aidelsburger},\ and\ \citenamefont {Bloch}}]{lohse2016}%
  \BibitemOpen
  \bibfield  {author} {\bibinfo {author} {\bibfnamefont {M.}~\bibnamefont
  {Lohse}}, \bibinfo {author} {\bibfnamefont {C.}~\bibnamefont {Schweizer}},
  \bibinfo {author} {\bibfnamefont {O.}~\bibnamefont {Zilberberg}}, \bibinfo
  {author} {\bibfnamefont {M.}~\bibnamefont {Aidelsburger}},\ and\ \bibinfo
  {author} {\bibfnamefont {I.}~\bibnamefont {Bloch}},\ }\bibfield  {title}
  {\bibinfo {title} {{A Thouless quantum pump with ultracold bosonic atoms in
  an optical superlattice}},\ }\href {https://doi.org/10.1038/nphys3584}
  {\bibfield  {journal} {\bibinfo  {journal} {Nature Physics}\ }\textbf
  {\bibinfo {volume} {12}},\ \bibinfo {pages} {350} (\bibinfo {year}
  {2016})}\BibitemShut {NoStop}%
\bibitem [{\citenamefont {Xie}\ \emph {et~al.}(2019)\citenamefont {Xie},
  \citenamefont {Gou}, \citenamefont {Xiao}, \citenamefont {Gadway},\ and\
  \citenamefont {Yan}}]{xie2019}%
  \BibitemOpen
  \bibfield  {author} {\bibinfo {author} {\bibfnamefont {D.}~\bibnamefont
  {Xie}}, \bibinfo {author} {\bibfnamefont {W.}~\bibnamefont {Gou}}, \bibinfo
  {author} {\bibfnamefont {T.}~\bibnamefont {Xiao}}, \bibinfo {author}
  {\bibfnamefont {B.}~\bibnamefont {Gadway}},\ and\ \bibinfo {author}
  {\bibfnamefont {B.}~\bibnamefont {Yan}},\ }\bibfield  {title} {\bibinfo
  {title} {{Topological characterizations of an extended Su--Schrieffer--Heeger
  model}},\ }\href {https://doi.org/10.1038/s41534-019-0159-6} {\bibfield
  {journal} {\bibinfo  {journal} {npj Quantum Information}\ }\textbf {\bibinfo
  {volume} {5}},\ \bibinfo {pages} {55} (\bibinfo {year} {2019})}\BibitemShut
  {NoStop}%
\bibitem [{\citenamefont {Mei}\ \emph {et~al.}(2018)\citenamefont {Mei},
  \citenamefont {Chen}, \citenamefont {Tian}, \citenamefont {Zhu},\ and\
  \citenamefont {Jia}}]{mei2018}%
  \BibitemOpen
  \bibfield  {author} {\bibinfo {author} {\bibfnamefont {F.}~\bibnamefont
  {Mei}}, \bibinfo {author} {\bibfnamefont {G.}~\bibnamefont {Chen}}, \bibinfo
  {author} {\bibfnamefont {L.}~\bibnamefont {Tian}}, \bibinfo {author}
  {\bibfnamefont {S.-L.}\ \bibnamefont {Zhu}},\ and\ \bibinfo {author}
  {\bibfnamefont {S.}~\bibnamefont {Jia}},\ }\bibfield  {title} {\bibinfo
  {title} {{Robust quantum state transfer via topological edge states in
  superconducting qubit chains}},\ }\href
  {https://doi.org/10.1103/PhysRevA.98.012331} {\bibfield  {journal} {\bibinfo
  {journal} {Physical Review A}\ }\textbf {\bibinfo {volume} {98}},\ \bibinfo
  {pages} {012331} (\bibinfo {year} {2018})}\BibitemShut {NoStop}%
\bibitem [{\citenamefont {Nie}\ and\ \citenamefont {Liu}(2020)}]{nie2020}%
  \BibitemOpen
  \bibfield  {author} {\bibinfo {author} {\bibfnamefont {W.}~\bibnamefont
  {Nie}}\ and\ \bibinfo {author} {\bibfnamefont {Y.-x.}\ \bibnamefont {Liu}},\
  }\bibfield  {title} {\bibinfo {title} {{Bandgap-assisted quantum control of
  topological edge states in a cavity}},\ }\href
  {https://doi.org/10.1103/PhysRevResearch.2.012076} {\bibfield  {journal}
  {\bibinfo  {journal} {Physical Review Research}\ }\textbf {\bibinfo {volume}
  {2}},\ \bibinfo {pages} {012076(R)} (\bibinfo {year} {2020})}\BibitemShut
  {NoStop}%
\bibitem [{\citenamefont {Guan}\ and\ \citenamefont
  {Chen}(2023)}]{guan2023interplay}%
  \BibitemOpen
  \bibfield  {author} {\bibinfo {author} {\bibfnamefont {X.}~\bibnamefont
  {Guan}}\ and\ \bibinfo {author} {\bibfnamefont {G.}~\bibnamefont {Chen}},\
  }\bibfield  {title} {\bibinfo {title} {{Interplay between topology and
  localization on superconducting circuits}},\ }\bibfield  {journal} {\bibinfo
  {journal} {arXiv preprint arXiv:2305.02486}\ }\href
  {https://doi.org/10.48550/arXiv.2305.02486} {10.48550/arXiv.2305.02486}
  (\bibinfo {year} {2023})\BibitemShut {NoStop}%
\bibitem [{\citenamefont {Petrescu}\ \emph {et~al.}(2018)\citenamefont
  {Petrescu}, \citenamefont {T\"{u}reci}, \citenamefont {Ustinov},\ and\
  \citenamefont {Pop}}]{PetrescuPRB2018}%
  \BibitemOpen
  \bibfield  {author} {\bibinfo {author} {\bibfnamefont {A.}~\bibnamefont
  {Petrescu}}, \bibinfo {author} {\bibfnamefont {H.~E.}\ \bibnamefont
  {T\"{u}reci}}, \bibinfo {author} {\bibfnamefont {A.~V.}\ \bibnamefont
  {Ustinov}},\ and\ \bibinfo {author} {\bibfnamefont {I.~M.}\ \bibnamefont
  {Pop}},\ }\bibfield  {title} {\bibinfo {title} {{Fluxon-based quantum
  simulation in circuit QED}},\ }\href
  {https://doi.org/10.1103/PhysRevB.98.174505} {\bibfield  {journal} {\bibinfo
  {journal} {Physical Review B}\ }\textbf {\bibinfo {volume} {98}},\ \bibinfo
  {pages} {174505} (\bibinfo {year} {2018})}\BibitemShut {NoStop}%
\bibitem [{\citenamefont {Parkin}\ \emph {et~al.}(2003)\citenamefont {Parkin},
  \citenamefont {Jiang}, \citenamefont {Kaiser}, \citenamefont {Panchula},
  \citenamefont {Roche},\ and\ \citenamefont {Samant}}]{ParkinIEEE2003}%
  \BibitemOpen
  \bibfield  {author} {\bibinfo {author} {\bibfnamefont {S.}~\bibnamefont
  {Parkin}}, \bibinfo {author} {\bibfnamefont {X.}~\bibnamefont {Jiang}},
  \bibinfo {author} {\bibfnamefont {C.}~\bibnamefont {Kaiser}}, \bibinfo
  {author} {\bibfnamefont {A.}~\bibnamefont {Panchula}}, \bibinfo {author}
  {\bibfnamefont {K.}~\bibnamefont {Roche}},\ and\ \bibinfo {author}
  {\bibfnamefont {M.}~\bibnamefont {Samant}},\ }\bibfield  {title} {\bibinfo
  {title} {{Magnetically engineered spintronic sensors and memory}},\ }\href
  {https://doi.org/10.1109/JPROC.2003.811807} {\bibfield  {journal} {\bibinfo
  {journal} {Proceedings of the IEEE}\ }\textbf {\bibinfo {volume} {91}},\
  \bibinfo {pages} {661} (\bibinfo {year} {2003})}\BibitemShut {NoStop}%
\bibitem [{\citenamefont {Parkin}\ \emph {et~al.}(2008)\citenamefont {Parkin},
  \citenamefont {Hayashi},\ and\ \citenamefont {Thomas}}]{ParkinScience2008}%
  \BibitemOpen
  \bibfield  {author} {\bibinfo {author} {\bibfnamefont {S.~S.~P.}\
  \bibnamefont {Parkin}}, \bibinfo {author} {\bibfnamefont {M.}~\bibnamefont
  {Hayashi}},\ and\ \bibinfo {author} {\bibfnamefont {L.}~\bibnamefont
  {Thomas}},\ }\bibfield  {title} {\bibinfo {title} {{Magnetic Domain-Wall
  Racetrack Memory}},\ }\href {https://doi.org/10.1126/science.1145799}
  {\bibfield  {journal} {\bibinfo  {journal} {Science}\ }\textbf {\bibinfo
  {volume} {320}},\ \bibinfo {pages} {190} (\bibinfo {year}
  {2008})}\BibitemShut {NoStop}%
\bibitem [{\citenamefont {Fert}\ \emph {et~al.}(2013)\citenamefont {Fert},
  \citenamefont {Cros},\ and\ \citenamefont {Sampaio}}]{FertNatNano2013}%
  \BibitemOpen
  \bibfield  {author} {\bibinfo {author} {\bibfnamefont {A.}~\bibnamefont
  {Fert}}, \bibinfo {author} {\bibfnamefont {V.}~\bibnamefont {Cros}},\ and\
  \bibinfo {author} {\bibfnamefont {J.~a.}\ \bibnamefont {Sampaio}},\
  }\bibfield  {title} {\bibinfo {title} {{Skyrmions on the track}},\ }\href
  {https://doi.org/10.1038/nnano.2013.29} {\bibfield  {journal} {\bibinfo
  {journal} {Nature Nanotechnology}\ }\textbf {\bibinfo {volume} {8}},\
  \bibinfo {pages} {152} (\bibinfo {year} {2013})}\BibitemShut {NoStop}%
\bibitem [{\citenamefont {Parkin}\ and\ \citenamefont
  {Yang}(2015)}]{ParkinNatNano2015}%
  \BibitemOpen
  \bibfield  {author} {\bibinfo {author} {\bibfnamefont {S.}~\bibnamefont
  {Parkin}}\ and\ \bibinfo {author} {\bibfnamefont {S.-H.}\ \bibnamefont
  {Yang}},\ }\bibfield  {title} {\bibinfo {title} {{Memory on the racetrack}},\
  }\href {https://doi.org/10.1038/nnano.2015.41} {\bibfield  {journal}
  {\bibinfo  {journal} {Nature Nanotechnology}\ }\textbf {\bibinfo {volume}
  {10}},\ \bibinfo {pages} {195} (\bibinfo {year} {2015})}\BibitemShut
  {NoStop}%
\bibitem [{\citenamefont {Vool}\ and\ \citenamefont
  {Devoret}(2017)}]{Vool_IJCTA2017}%
  \BibitemOpen
  \bibfield  {author} {\bibinfo {author} {\bibfnamefont {U.}~\bibnamefont
  {Vool}}\ and\ \bibinfo {author} {\bibfnamefont {M.}~\bibnamefont {Devoret}},\
  }\bibfield  {title} {\bibinfo {title} {{Introduction to quantum
  electromagnetic circuits}},\ }\href
  {https://doi.org/https://doi.org/10.1002/cta.2359} {\bibfield  {journal}
  {\bibinfo  {journal} {International Journal of Circuit Theory and
  Applications}\ }\textbf {\bibinfo {volume} {45}},\ \bibinfo {pages} {897}
  (\bibinfo {year} {2017})}\BibitemShut {NoStop}%
\bibitem [{\citenamefont {Fazio}\ and\ \citenamefont {{van der
  Zant}}(2001)}]{FazioPhysRep2001}%
  \BibitemOpen
  \bibfield  {author} {\bibinfo {author} {\bibfnamefont {R.}~\bibnamefont
  {Fazio}}\ and\ \bibinfo {author} {\bibfnamefont {H.}~\bibnamefont {{van der
  Zant}}},\ }\bibfield  {title} {\bibinfo {title} {Quantum phase transitions
  and vortex dynamics in superconducting networks},\ }\href
  {https://doi.org/https://doi.org/10.1016/S0370-1573(01)00022-9} {\bibfield
  {journal} {\bibinfo  {journal} {Physics Reports}\ }\textbf {\bibinfo {volume}
  {355}},\ \bibinfo {pages} {235} (\bibinfo {year} {2001})}\BibitemShut
  {NoStop}%
\bibitem [{Note1()}]{Note1}%
  \BibitemOpen
  \bibinfo {note} {The mean-field contribution of the charging energy is an
  extra tight-binding element $\DOTSB \sum@ \slimits@ _{i, j} m_{i, j} \protect
  \, b^{\dagger }_{i} \protect \, b_{j}$, where $m_{i, j} = m^{\ast }_{j, i}$
  is determined self-consistently \protect \[ m_{i, j} = \protect \tilde
  {C}^{-1}_{i, j} \protect \, \expval {b^{\dagger }_{j} \protect \, b_{i}}.
  \protect \] The assumption is that this matrix is a small correction to the
  hopping elements already included in Eq.~(\ref {eq:HopHam1}).}\BibitemShut
  {Stop}%
\bibitem [{\citenamefont {Zak}(1989)}]{zak1989}%
  \BibitemOpen
  \bibfield  {author} {\bibinfo {author} {\bibfnamefont {J.}~\bibnamefont
  {Zak}},\ }\bibfield  {title} {\bibinfo {title} {{Berry's phase for energy
  bands in solids}},\ }\href {https://doi.org/10.1103/PhysRevLett.62.2747}
  {\bibfield  {journal} {\bibinfo  {journal} {Physical Review Letters}\
  }\textbf {\bibinfo {volume} {62}},\ \bibinfo {pages} {2747} (\bibinfo {year}
  {1989})}\BibitemShut {NoStop}%
\bibitem [{\citenamefont {Eliashvili}\ \emph {et~al.}(2017)\citenamefont
  {Eliashvili}, \citenamefont {Kereselidze}, \citenamefont {Tsitsishvili},\
  and\ \citenamefont {Tsitsishvili}}]{Eliashvili2017}%
  \BibitemOpen
  \bibfield  {author} {\bibinfo {author} {\bibfnamefont {M.}~\bibnamefont
  {Eliashvili}}, \bibinfo {author} {\bibfnamefont {D.}~\bibnamefont
  {Kereselidze}}, \bibinfo {author} {\bibfnamefont {G.}~\bibnamefont
  {Tsitsishvili}},\ and\ \bibinfo {author} {\bibfnamefont {M.}~\bibnamefont
  {Tsitsishvili}},\ }\bibfield  {title} {\bibinfo {title} {{Edge States of a
  Periodic Chain with Four-Band Energy Spectrum}},\ }\href
  {https://doi.org/10.7566/JPSJ.86.074712} {\bibfield  {journal} {\bibinfo
  {journal} {Journal of the Physical Society of Japan}\ }\textbf {\bibinfo
  {volume} {86}},\ \bibinfo {pages} {074712} (\bibinfo {year}
  {2017})}\BibitemShut {NoStop}%
\bibitem [{\citenamefont {Maffei}\ \emph {et~al.}(2018)\citenamefont {Maffei},
  \citenamefont {Dauphin}, \citenamefont {Cardano}, \citenamefont
  {Lewenstein},\ and\ \citenamefont {Massignan}}]{Maffei2018}%
  \BibitemOpen
  \bibfield  {author} {\bibinfo {author} {\bibfnamefont {M.}~\bibnamefont
  {Maffei}}, \bibinfo {author} {\bibfnamefont {A.}~\bibnamefont {Dauphin}},
  \bibinfo {author} {\bibfnamefont {F.}~\bibnamefont {Cardano}}, \bibinfo
  {author} {\bibfnamefont {M.}~\bibnamefont {Lewenstein}},\ and\ \bibinfo
  {author} {\bibfnamefont {P.}~\bibnamefont {Massignan}},\ }\bibfield  {title}
  {\bibinfo {title} {{Topological characterization of chiral models through
  their long time dynamics}},\ }\href
  {https://doi.org/10.1088/1367-2630/aa9d4c} {\bibfield  {journal} {\bibinfo
  {journal} {New Journal of Physics}\ }\textbf {\bibinfo {volume} {20}},\
  \bibinfo {pages} {013023} (\bibinfo {year} {2018})}\BibitemShut {NoStop}%
\bibitem [{\citenamefont {Wong}(2022)}]{wong2022}%
  \BibitemOpen
  \bibfield  {author} {\bibinfo {author} {\bibfnamefont {P.~J.}\ \bibnamefont
  {Wong}},\ }\emph {\bibinfo {title} {{Interactions and Topology in Quantum
  Matter: Auxiliary Field Approach \& Generalized SSH Models}}},\ \href@noop {}
  {Ph.D. thesis},\ \bibinfo  {school} {University College Dublin} (\bibinfo
  {year} {2022})\BibitemShut {NoStop}%
\bibitem [{\citenamefont {Fukui}\ \emph {et~al.}(2005)\citenamefont {Fukui},
  \citenamefont {Hatsugai},\ and\ \citenamefont {Suzuki}}]{fukui2005}%
  \BibitemOpen
  \bibfield  {author} {\bibinfo {author} {\bibfnamefont {T.}~\bibnamefont
  {Fukui}}, \bibinfo {author} {\bibfnamefont {Y.}~\bibnamefont {Hatsugai}},\
  and\ \bibinfo {author} {\bibfnamefont {H.}~\bibnamefont {Suzuki}},\
  }\bibfield  {title} {\bibinfo {title} {{Chern Numbers in Discretized
  Brillouin Zone: Efficient Method of Computing (Spin) Hall Conductances}},\
  }\href {https://doi.org/10.1143/JPSJ.74.1674} {\bibfield  {journal} {\bibinfo
   {journal} {Journal of the Physical Society of Japan}\ }\textbf {\bibinfo
  {volume} {74}},\ \bibinfo {pages} {1674} (\bibinfo {year}
  {2005})}\BibitemShut {NoStop}%
\bibitem [{\citenamefont {Seoane~Souto}\ \emph {et~al.}(2020)\citenamefont
  {Seoane~Souto}, \citenamefont {Kuzmanovski},\ and\ \citenamefont
  {Balatsky}}]{SoutoPRR2020}%
  \BibitemOpen
  \bibfield  {author} {\bibinfo {author} {\bibfnamefont {R.}~\bibnamefont
  {Seoane~Souto}}, \bibinfo {author} {\bibfnamefont {D.}~\bibnamefont
  {Kuzmanovski}},\ and\ \bibinfo {author} {\bibfnamefont {A.~V.}\ \bibnamefont
  {Balatsky}},\ }\bibfield  {title} {\bibinfo {title} {{Signatures of
  odd-frequency pairing in the Josephson junction current noise}},\ }\href
  {https://doi.org/10.1103/PhysRevResearch.2.043193} {\bibfield  {journal}
  {\bibinfo  {journal} {Physical Review Research}\ }\textbf {\bibinfo {volume}
  {2}},\ \bibinfo {pages} {043193} (\bibinfo {year} {2020})}\BibitemShut
  {NoStop}%
\end{thebibliography}%

\end{document}